\documentclass[pre,aps,10pt,superscriptaddress,twocolumn,floatfix]{revtex4-1}

\usepackage[nice]{nicefrac}
\usepackage{graphicx,color}
\usepackage{amsmath,amssymb,bm,bbm}
\usepackage{amsmath}
\usepackage{braket}
\usepackage{bbold}
\usepackage{float}

\usepackage[plainpages=false,pdfpagelabels,colorlinks=true,linkcolor=red,urlcolor=blue,citecolor=blue,pdftitle={},pdfauthor={},pdfdisplaydoctitle=true,pdfduplex=DuplexFlipLongEdge]{hyperref}

\definecolor{darkred}{rgb}{0.90,0.2,0.2}
\definecolor{darkgreen}{rgb}{0,0.60,.2}
\definecolor{darkblue}{rgb}{0.1,0.3,1}
\definecolor{grey}{cmyk}{0,0,0,0.25}
\definecolor{orange}{cmyk}{0,0.6,0.8,0}

\begin{document}

\title{Critical Dynamics in Short-Range Quadratic Hamiltonians}

\author{Miroslav Hopjan}
\affiliation{Department of Theoretical Physics, J. Stefan Institute, SI-1000 Ljubljana, Slovenia}
\author{Lev Vidmar}
\affiliation{Department of Theoretical Physics, J. Stefan Institute, SI-1000 Ljubljana, Slovenia}
\affiliation{Department of Physics, Faculty of Mathematics and Physics, University of Ljubljana, SI-1000 Ljubljana, Slovenia\looseness=-1}

\begin{abstract}
We investigate critical transport and the dynamical exponent through the spreading of an initially localized particle in quadratic Hamiltonians with short-range hopping in lattice dimension $d_l$.
We consider critical dynamics that emerges when the Thouless time, i.e., the saturation time of the mean-squared displacement, approaches the typical Heisenberg time.
We establish a relation, $z=d_l/d_s$, linking the critical dynamical exponent $z$ to $d_l$ and to the spectral fractal dimension $d_s$.
This result has notable implications: it says that superdiffusive transport in $d_l\geq 2$ and diffusive transport in $d_l\geq 3$ cannot be critical in the sense defined above.
Our findings clarify previous results on disordered and quasiperiodic models and, through Fibonacci potential models in two and three dimensions, provide non-trivial examples of critical dynamics in systems with $d_l\neq1$ and $d_s\neq1$.
\end{abstract}

\maketitle

\textit{Introduction.}---The transport of mass, energy, and other conserved quantities is a fundamental characteristic of physical systems~\cite{Datta_1995,Ziman01,Adams03,Oberhofer17}, often described by the diffusion equation~\cite{Mehrer07,Bokstein2018}.
At the nanoscale, where quantum effects become significant, classical transport theories must be adjusted to account for quantum phenomena~\cite{Bonitz98,Thoss18,Bertini21,waintal2024}.
Interestingly, diffusion remains relevant even in quantum transport~\cite{Steinigeweg14, Karrasch14, varma2019diffusive,Gopalakrishnan19, DeNardis19, Richter19, Wurtz20, Schubert21, Bertini21, Prelovsek21, Prelovsek22, Nandy23, Prelovsek23b, Wang24, kraft2024, Ampelogiannis2025}; however, there exist notable exceptions.
For instance, Google’s recent quantum simulator experiment~\cite{prosen24_short} explored superdiffusive transport in integrable models~\cite{Znidaric11,Ljubotina17,Ilievski18,Ljubotina19,DeNardis20,Scheie21,Bulchandani21,  Ilievski21,Wei22,DeNardis23,Krajnik24,Gopalakrishnan24,Bastianello24} and their relation to the Kardar-Parisi-Zhang universality~\cite{Kardar86}. Another prominent example is slow subdiffusive dynamics in disordered~\cite{Prelovsek17,luitz_barlev_17, Sierant_2025}~ and quasi-periodic~\cite{luitz_barlev_17, Sierant_2025, Chiaracane20} systems. 
Thus, the diffusion often serves as the reference point in transport studies, while its absence often signals critical behavior, such as integrability~\cite{Caux_2011,Calabrese_2016} or localization~\cite{anderson_58}. 
However, not all cases of non-diffusive transport indicate criticality; some may simply result from long-lived transient effects. Understanding the mechanisms underlying critical dynamics is therefore essential.

\begin{figure}[b]
\centering
\includegraphics[width=0.98\columnwidth]{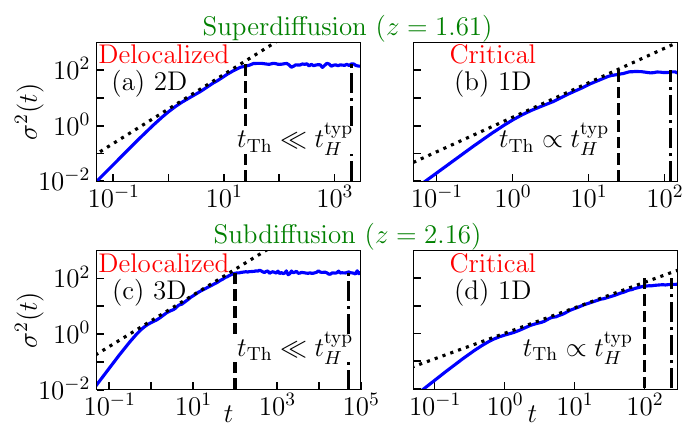}
\caption{
The dynamics of the mean-squared displacement $\sigma^2(t)$, see Eq.~\eqref{def_sigma2}, in the separable Fibonacci models in dimensions one (1D), two (2D) and three (3D).
We set $L=40$ and the potentials (a-b) $h=2$, (c-d) $h=4$, see Eq.~\eqref{eq:ham} and the text below.
Dotted lines, which mark the slope of $\sigma^2(t)$, suggests that the dynamical exponent $z$ is superdiffusive ($1<z<2$) in (a-b) and subdiffusive ($z>2$) in (c-d).
In (b) and (d), the dynamics are critical since they fulfill Eq.~\eqref{critical}, while in (a) and (c) they are not.
}
\label{fig_example}
\end{figure}

In systems governed by quadratic Hamiltonians, the dynamics are often characterized by the spreading of a local excitation~\cite{Bertini21} through the mean-squared displacement $\sigma^2(t)$~\cite{Frenkel02,Troisi06, Troisi11, Wang11, Chen17a, Kloss19, tenBrink22,  Chen17a, Khan21, Birschitzky24,Marquardt21,Schirripa24,Bindech24,Bertini21} and the associated dynamical exponent $z$ (defined in Eqs.~\eqref{def_sigma}-\eqref{def_sigma2} below).
Despite extensive studies on quadratic systems, several key questions remain unresolved:
How should critical dynamics be defined, and is there a simple relation for the dynamical exponent $z$?
When do non-diffusive transport types, such as superdiffusion and subdiffusion, indicate critical dynamics rather than transient effects? 
Figure~\ref{fig_example} highlights this distinction in Fibonacci models across one to three dimensions.
For a fixed $z$, non-trivial dynamics may be considered critical since $\sigma^2(t)$ saturates near the system's longest time scale, the typical Heisenberg time $t_H^{\rm typ}$ [Figs.~\ref{fig_example}(b) and~\ref{fig_example}(d)].
Alternatively, they may be viewed as transient effects if the dynamics lead to complete
wave-packet delocalization at times much shorter than $t_H^{\rm typ}$ [Figs.~\ref{fig_example}(a) and~\ref{fig_example}(c)].

In this Letter, we revisit the concept of critical transport. 
We examine the critical dynamics that satisfies the following condition:
the saturation time of the wave-packet mean-squared displacement, referred to as the Thouless time $t_{\rm Th}$, scales with the linear system size $L$ in the same way as the typical Heisenberg time $t_{H}^{\rm{typ}}$, i.e., the inverse typical level spacing,
\begin{equation}\label{critical}
    t_{\rm Th}(L) \propto t_{H}^{\rm{typ}}(L)\;.
\end{equation}
We then establish a simple relation for the dynamical exponent at criticality:
\begin{equation} \label{def_z_main}
    z = \frac{d_l}{d_s}\;,
\end{equation}
where $d_l$ is the lattice dimension and $d_s$ is the spectral fractal dimension.
Consequently, if one uses Eq.~\eqref{critical} as the definition of critical dynamics, superdiffusive transport ($1<z<2$) cannot be critical in dimensions $d_l \geq 2$, and diffusive transport ($z=2$) cannot be critical in $d_l \geq 3$.
We provide a simple but non-trivial example in which these predictions can be observed.

\textit{Preliminaries.}---We consider systems described by quadratic fermionic Hamiltonians, i.e., by bilinear forms in creation and annihilation operators.
The dynamics of a particle, or an excitation, are then governed by a lattice potential, and the mean-squared displacement  $\sigma^2_H(t)$ can be defined within a single-particle space. For a particle initially localized at site ${\bm i_0}=(i_{0}^{1},\dots,i_{0}^{d_l}) $ of a $d_l$-dimensional lattice, $\sigma^2_H(t)$ is defined as
\begin{equation} \label{def_sigma}
    \sigma^2_H(t) = \sum_{{\bm i}} |{\bm i}-{\bm i}_0|^2 |\langle {\bm i}|e^{-i\hat{H}t}|{\bm i_0} \rangle|^2 \;,
\end{equation}
where $|{\bm i}-{\bm i}_0|$ is the Euclidean distance between the initial site ${\bm i_0}$ and site ${\bm i}$, 
the sum runs over all lattice sites ${\bm i}$, and the time evolution is governed by the time-independent Hamiltonian $\hat{H}$. 
As we typically consider an ensemble of Hamiltonians, where for each realization we set the particle in the center of the lattice, we also define the Hamiltonian-averaged mean-squared displacement $\sigma^2(t)$.
The time-evolution of the latter is then used to define the dynamical exponent $z$,
\begin{equation} \label{def_sigma2}
    \sigma^2(t) = \langle \sigma^2_H(t) \rangle_H \propto t^{2/z}\;,
\end{equation}
where $\langle ... \rangle_H$ denotes the average over Hamiltonian realizations.

We note that $z$ in Eq.~\eqref{def_sigma2} is defined within a single-particle set-up.
The same $z$ can be also measured from the dynamics of many-body states, e.g., the particle imbalance of the domain wall state~\cite{varma2019diffusive}, the entanglement entropy~\cite{luitz_barlev_17}, or the surface roughness~\cite{Fujimoto21,Bhakuni24,Sreemayee24}.

The numerical results in this Letter are obtained for the Fibonacci models in 1D, 2D and 3D lattices, defined as
\begin{equation}
\label{eq:ham}
\hat H= -{J}\sum_{\langle {\bm i}{\bm j}\rangle}^{} (\hat{c}_{{\bm i}}^{\dagger}\hat{c}_{{\bm j}}^{}+ \hat c_{\bm j}^\dagger \hat c_{\bm i}) + \sum_{{\bm i}=1}^{V}\epsilon_{{\bm i}}\hat{n}_{{\bm i}}^{}\;,
\end{equation}
where $\hat{c}_{{\bm j}}^{\dagger}$ ($\hat{c}_{{\bm j}}^{}$) are the fermionic creation (annihilation) operators at site ${\bm j}$, $J\equiv 1$ is the hopping matrix element between nearest neighbor sites,
$\hat{n}_{{\bm i}}^{}=\hat{c}_{{\bm i}}^{\dagger}\hat{c}_{{\bm i}}^{}$ is the site occupation operator, and $\epsilon_{{\bm i}}$ is the on-site potential.
The potential is separable and can be written as a sum, 
$\epsilon_{{\bm i}}=\sum_{l=1}^{d_l} \epsilon_{i^{l}}$,
where $\epsilon_{i^{l}}^{}$ is the potential value in direction $l$, with $i^{l} \in \{1,\dots,L\}$.
The potential values in each direction are chosen by considering a randomly chosen sub-sequence of length $L$ from the Fibonacci sequence of large length, $L_F=10^{6}$. The Fibonacci sequence $\epsilon_{f}$ for $f \in \{1,\dots,L_F\}$ is generated from $\epsilon_{f}=2hV(fg)-h$, where $g=(\sqrt{5}-1)/2$ and $V(x)=[x+g]-[x]$, with $[x]$ being integer part of $x$~\cite{varma2019diffusive}.
For $d_l=1$, the system reduces to the usual 1D Fibonacci model~\cite{Ketzmerick92, Ketzmerick97, varma2019diffusive}. 

As a technical remark, we note that the advantage of the separability of potentials in 2D (square lattice) and 3D (cubic lattice) is that the eigenenergy spectrum consists of a sum of eigenenergies of Fibonacci chains in each directions~\cite{Thiem13a}, see End Matter for details.
This allows us to reach spectra of systems of large linear system sizes, $L=10000$ in 2D and $L=500$ in 3D.

\textit{Characteristic times.}---We next define the two characteristic time scales relevant for our study, the Thouless and the Heisenberg time.
The Thouless time $t_{\rm Th}$ is defined as the time when $\sigma^2(t)$ saturates to a constant value, i.e., the time at which the particle reaches the lattice boundary.
We mark $t_{\rm Th}$ by the vertical dashed lines in Fig.~\ref{fig_example}.
For the Heisenberg time, we focus on its typical value.
The typical Heisenberg time is defined as the inverse of the typical eigenvalue spacing times $2\pi$,
\begin{equation}\label{Heis_typ_def}
    t_{H}^{\rm{typ}}= \frac{2\pi}{\delta E^{\rm{typ}}}=\frac{2\pi}{\exp(\langle{\ln({E_\mu-E_{\mu-1}})}\rangle_H)}\;,
\end{equation}
where $E_\mu$ are the eigenvalues of Hamiltonian $\hat{H}$, $\hat{H}|\mu \rangle=E_\mu|\mu \rangle$, with $\mu=1,...,V$ and $V=L^{d_l}$ is the total number of lattice sites.
In Eq.~\eqref{Heis_typ_def} we only consider nonzero energy gaps, $E_\mu-E_{\mu-1}\neq 0$, i.e., degeneracies (when exist) are excluded as they do not contribute non-trivially to the time evolution of the system. 
We mark $t_{H}^{\rm{typ}}$ by the vertical dashed-dotted lines in Fig.~\ref{fig_example}.

While the Thouless time is usually interpreted as the longest physically relevant time scale, the Heisenberg time sets the upper bound for the accessible time scales in a finite system.
Intuitively, the longest time scale in the dynamics of finite systems is associated with the small gap values, and hence we consider the typical rather than the average Heisenberg time.
In quantum-chaotic systems and in localized systems, one usually encounters $t_{H}^{\rm{typ}} \propto V$ and hence the typical level spacing is proportional to the average level spacing.
However, for systems affine to clustering of eigenvalues, $t_{H}^{\rm{typ}}$ may scale with a higher power of $V$.
We parametrize this behavior as
\begin{equation}\label{Heis_scaling}
    t_{H}^{\rm{typ}} = c_H V^{n}\;,
\end{equation}
where $c_H$ is a constant and $n$ is a number that characterizes the level clustering.
Hence, for $n>1$ the timescale of non-trivial dynamics becomes significantly longer than in systems with $n=1$.

\begin{figure}[t!]
\centering
\includegraphics[width=0.98\columnwidth]{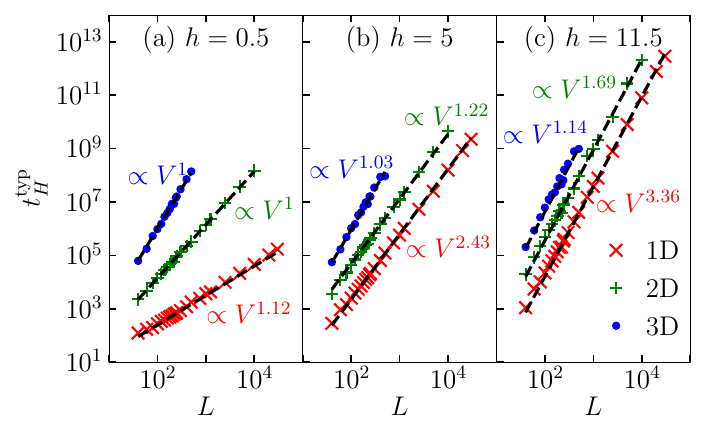}
\caption{
Scaling of the typical Heisenberg time, $t_{H}^{\rm{typ}}$, with linear systems size $L$, in the separable 1D, 2D, and 3D Fibonacci models.
Results are shown for potentials (a) $h=0.5$, (b) $h=5.0$ and (c) and $h=11.5$.
Lines are fits to Eq.~\eqref{Heis_scaling}, with the values of $n$ given in each panel, and $V=L^{d_l}$ is the total number of lattice sites.
}
\label{fig_n}
\end{figure}

A known example for $n\neq 1$ in Eq.~\eqref{Heis_scaling} is the one-dimensional (1D) Aubry-André model, for which $n=2$ at criticality~\cite{hopjan2023}.
Another example are the separable 1D, 2D and 3D Fibonacci models, for which the scaling of $t_{H}^{\rm{typ}}$ versus $L$ are shown in Fig.~\ref{fig_n}.
We find $n=1$ for the 2D and 3D systems at potential $h=0.5$, see Fig.~\ref{fig_n}(a), as well as for the 3D system at potential $h=5$, see Fig.~\ref{fig_n}(b).
All other cases displayed in Fig.~\ref{fig_n} exhibit $n>1$ as consequence of the level clustering in the spectrum. 

\begin{figure}[!t]
\centering
\includegraphics[width=0.98\columnwidth]{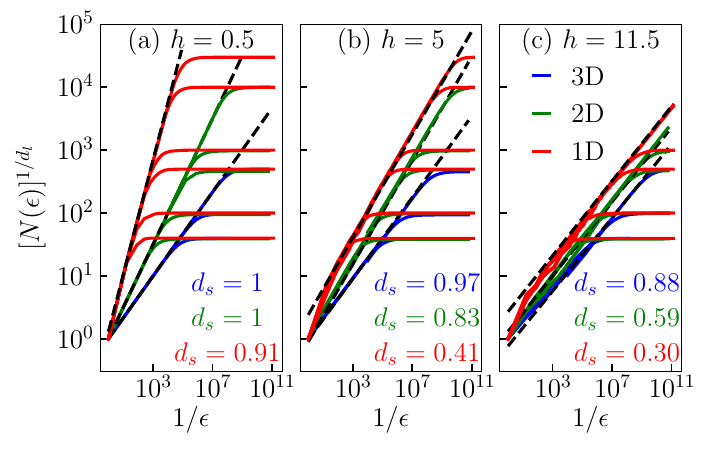}
\caption{Scaling of $[N(\epsilon)]^{1/d_l}$ with the inverse box length $1/\epsilon$, see Eq.~\eqref{N_scaling}, in separable Fibonacci models of various dimension for (a) $h=0.5$, (b) $h=5.0$ and (c) and $h=11.5$. The linear sizes are $L=40, 100, 500, 1000, 10000, 30000$ for 1D lattice, $L=40, 100, 500, 1000, 10000$ for 2D lattice and $L=40, 100, 500, 1000$ for the 3D lattice. The slopes of $[N(\epsilon)]^{1/d_l}$, given by the ratio ${d_s/d_l}$, are extracted from the fits (dashed lines). The extracted values of $d_s$ should be compared to the values of $n$ in Fig.~\ref{fig_n}, where the same model parameters were used. The numerical results confirm validity of the conjecture in Eq.~\eqref{n_ds} to a second digit.}
\label{fig_ds}
\end{figure}

\textit{Spectral fractality.}---We now show that the exponent $n$ from Eq.~\eqref{Heis_scaling} is connected to the fractal dimension $d_s$ of the Hamiltonian spectrum~\cite{Kohmoto83, Kohmoto84, Tang86, Halsey86, Kohmoto87}. 
We extract the latter using the box counting method.
To this end, we transform the eigenenergies $E_\mu$ as $\epsilon_\mu=(E_\mu-E^{\rm min}_\mu)/(E^{\rm max}_\mu-E^{\rm min}_\mu)$, where $E^{\rm max}_\mu$ is the maximal energy and $E^{\rm min}_\mu$ is the minimal energy, such that the transformed eigenspectrum $\epsilon_\mu$ spans the interval $[0,1]$. The latter is then divided into $1/\epsilon$ boxes $B$ of length $\epsilon$. For each Hamiltonian spectrum we define the scaling function of moment $q$,
\begin{equation}\label{N_q}
     N_{s,H}^{(q)}(\epsilon) = \sum_{B\neq B_\varnothing} (\sum_{\epsilon_\mu \in B} 1)^{q} \;,
\end{equation}
where the first sum runs over non-empty boxes, $B\neq B_\varnothing$, and the second sum counts the number of levels in each box.
The scaling function at $q=0$ can be interpreted as the number of boxes that contain at least one of the eigenenergies $\epsilon_\mu$.
The averaged scaling function,
\begin{equation}\label{N_q_scaling}
      N_{s}^{(q)}(\epsilon) =\Big\langle N_{s,H}^{(q)}(\epsilon)\Big\rangle_H \propto \epsilon^{(q-1)d_s^{(q)}},
\end{equation}
defines the spectral fractal dimension $d_s^{(q)}$.
Here, we mainly focus on the scaling function at $q=0$, $N(\epsilon)\equiv N_{s}^{(0)}(\epsilon)$, as it is connected to the typical Heisenberg time, see the discussion below.
Its scaling can be expressed as 
\begin{equation}\label{N_scaling}
     N(\epsilon)= c_N\Bigl(\frac{1}{\epsilon}\Bigr)^{d_s}\;,
\end{equation}
where $c_N$ is a constant and $d_s\equiv d_s^{(0)}$. 
We numerically test Eq.~\eqref{N_scaling} in Fig.$~$\ref{fig_ds} of End Matter, and we show that it indeed represents a meaningful ansatz to extract the spectral fractal dimension $d_s$.

Now, we argue that the exponents $n$ in Eq.~\eqref{Heis_scaling} and $d_s$ in Eq.~\eqref{N_scaling} are related as
\begin{equation}\label{n_ds}
    n = \frac{1}{d_s}\;.
\end{equation}
The argument for validity of Eq.~\eqref{n_ds} goes as follows.
Since the typical level spacing $\delta E^{\rm typ}$ is related to its median, one expects that at $\epsilon = \delta E^{\rm typ} \propto (t_H^{\rm typ})^{-1}$, the number of occupied boxes $N$ is proportional to $V$ [one may approximate it as $N(\epsilon=\delta E^{\rm typ}) = V/m$, with $m=O(1)$].
Then, it follows from Eq.~\eqref{N_scaling} that $V^{1/d_s} \propto t_H^{\rm typ}$, which is identical to Eq.~\eqref{Heis_scaling} assuming the relation $n=1/d_s$ in Eq.~\eqref{n_ds}.
We also tested Eq.~\eqref{n_ds} numerically in Figs.$~$\ref{fig_n} and$~$\ref{fig_ds}, in which we display the values of $n$ and $1/d_s$, respectively, for the same system parameters.
We find excellent agreement between the numerical values of $n$ and $1/d_s$.

{\it Derivation of critical dynamical exponent.}---We now turn to the derivation of the critical dynamical exponent $z$ from Eq.~\eqref{def_z_main}.
In the first step, we show that the Thouless time $t_{\rm Th}$ generally depends on the dynamical exponent $z$.
For systems that are not localized (i.e., at $z<\infty$), the mean displacement $\sigma(t)$ approaches the upper bound $\overline{\sigma}$ in the infinite-time limit, which is proportional to the linear system size,
\begin{equation}\label{sigma_inf}
    \overline{\sigma} \equiv \lim_{t \rightarrow \infty} \sigma(t) \propto L\;.
\end{equation}
The upper bound $\overline{\sigma}$ corresponds to the plateau of $\sigma(t)$ in Fig.~\ref{fig_example}.
We parameterize the Thouless time, i.e., the time at which the plateau $\overline{\sigma}$ is reached, as $t_{\rm Th} \propto L^\alpha$, and the mean particle displacement, cf.~Eq.~\eqref{def_sigma2}, as $\sigma(t) = c_\sigma t^{1/z}$, where $c_\sigma$ is a constant.
Then, requiring that $\sigma(t=t_{\rm Th}) = c_\sigma t_{\rm Th}^{1/z} \propto L$, one gets $a=z$ and hence
\begin{equation}\label{Thouless}
    t_{\rm Th} \propto L^z \;.
\end{equation}
This is an expected result and it is consistent, e.g., with ballistic dynamics at $z=1$ and diffusion at $z=2$.

The scaling of the typical Heisenberg time is, in contrast, governed by the spectral properties.
Combining Eq.~\eqref{Heis_scaling} with Eq.~\eqref{n_ds}, its scaling with the linear system size $L$ can be expressed as
\begin{equation}\label{def_Heis_L}
   t_{H}^{\rm typ} \propto V^{\frac{1}{d_s}} \propto L^{\frac{d_l}{d_s}} \;.
\end{equation}
The two seemingly independent results in Eqs.~\eqref{Thouless} and~\eqref{def_Heis_L} can be related at criticality when both times exhibit an identical dependence on $L$, as suggested by Eq.~\eqref{critical}.
Hence, the requirement for the scaling of the Thouless time to match the scaling of the typical Heisenberg time gives rise to the relation $z=d_l/d_s$ in Eq.~\eqref{def_z_main}, which is the main result of this Letter.

We note that our notion of critical slowing down relies on the system's tendency to localize in real space.
If localization is expected to occur in other physically relevant spaces, such as quasi-momentum space, the definition of Thouless time must be adjusted accordingly.

A remarkable consequence of Eq.~\eqref{def_z_main} is that it provides bounds on the critical dynamical exponent $z$, if one uses Eq.~\eqref{critical} as the definition of critical dynamics.
Since we expect the fractal dimension $d_s$ to be limited to the interval $d_s \in [0,1]$, it follows that $z \geq d_l$.
One can then argue that $z^* = d_l$ represents the fastest critical dynamics.
Hence, ballistic transport can only be critical in 1D, superdiffusive transport cannot be critical in dimensions two or higher, and diffusive transport cannot be critical in dimensions higher than two.

We note that many previous works attempted for establishing a general connection between the dynamical exponent $z$ and different versions of spectral fractal dimension, see, e.g., Refs.~\cite{Abe87, Hiramoto88a, Hiramoto88b, Guarneri89, Geisel91, Geisel91b, Geisel1992, Lima91, Artuso92a, Artuso92b, Guarneri93, Evangelou93, Guarneri94, Wilkinson94, Fleischmann95, Guarneri95, Zhong95, Kawarabayashi95, Picheon96, Brandes96, Ketzmerick97, Huckestein97, Mantica97, Huckestein98, Huckestein99, Kawarabayashi99, Guarneri99, Lillo00, Killip01, Yuan00, Zhong00, Zhong01, Guarneri02, Cerovski05, Damanik06, Jitomirskaya07, Ng07, Thiem09, Schreiber09, Thiem10, Thiem12, Zhang12, Thiem13, Thiem13a, Shamis23}.
Our study suggests that the simple expression for $z$, cf.~Eq.~\eqref{def_z_main}, indeed exists.
However, it is limited to the regime of critical dynamics, which is given by a rather stringent criterion from Eq.~\eqref{critical}.

Another notable consequence of the critical dynamics defined via Eq.~\eqref{critical} is the emergence of scale invariant principle~\cite{hopjan2023,Hopjan23b,Jiricek24,Hopjan24}.
The latter is obtained upon rescaling the time as $t \rightarrow \tau={t}/{t_{H}^{\rm{typ}}}$, i.e., the dynamics of $\sigma$ becomes scale-invariant in $\tau$ when rescaled as $\sigma \rightarrow \sigma'={\sigma}/{L}$. The scale-invariance of $\sigma'$ in $\tau$ is similar to the scale-invariance of the spectral form factor~\cite{suntajs_prosen_21}, the survival probability~\cite{hopjan2023,Hopjan23b} and simple observables such as imbalance upon the corresponding rescaling in y-axis~\cite{Jiricek24,Hopjan24}.
All these quantities are scale-invariant both for $\tau<1$ and $\tau>1$, which hints on generality of the emergent scale invariance at criticality.

This should be contrasted to rescaling $t \rightarrow \tau'={t}/{t_{\rm Th}}$ and $ \sigma \rightarrow \sigma'={\sigma}/{L}$, known as Family-Vicsek scaling law~\cite{Vicsek84}, see, e.g., Ref.~\cite{Bhakuni24}. The dynamics of $\sigma'$ is indeed scale invariant, both before $\tau'<1$ and after $\tau'>1$ the Thouless time. 
However, since the collapse of $\sigma'$
in $\tau'$ is present both for the delocalized and critical dynamics, it cannot be used to detect criticality. Also, such time rescaling appears to be relevant for $\sigma$ and related quantities, and not for other measures of the dynamics such as the spectral form factor. 

\textit{Numerical examples.}---We now test our main result, Eq.~\eqref{def_z_main}, for the paradigmatic disordered and quasi-periodic quadratic models.
We first argue that Eq.~\eqref{def_z_main} is consistent with the available results in the literature.
Perhaps the most studied disordered model is the Anderson model~\cite{anderson_58}, for which $d_s=1$ at criticality~\cite{Ketzmerick97}.
It was found that $z=3$ in a 3D lattice~\cite{Ohtsuki1997,sierant_delande_20} and $z=5$ in a 5D lattice~\cite{sierant_delande_20}, consistent with Eq.~\eqref{def_z_main}.
For the quasi-periodic models, most studies focused on the 1D Aubry-André model and the 1D Fibonacci model, i.e., at $d_l=1$.
In the former model, it was found that $1/z=d_s=0.5$ at the critical point~\cite{Geisel91, Ketzmerick92, Ketzmerick97}.
The latter model, which we reinvestigate in Fig.~\ref{fig_z} below, is critical for any potential strength $h$ such that $1/z=d_s\in [0,1]$ is a continuously varying function of $h$~\cite{Ketzmerick92,Ketzmerick97}.
Both results are in agreement with Eq.~\eqref{def_z_main}.

\begin{figure}[t!]
\centering
\includegraphics[width=0.98\columnwidth]{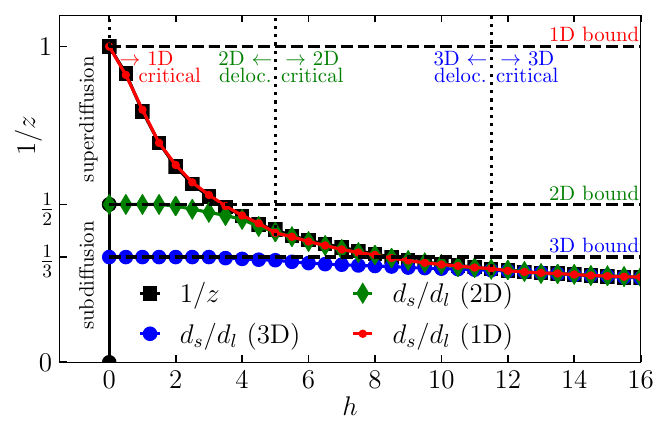}
\caption{Inverse dynamical exponent $1/z$ and the ratio $d_s/d_l$ of spectral fractal dimension $d_s$ and the lattice dimension $d_l$ as a function of the potential strength in 1D, 2D and 3D separable Fibonacci models.
}
\label{fig_z}
\end{figure}

To our knowledge, all previous studies of critical dynamics considered either $d_l=1$ and $d_s\neq 1$, or $d_l\neq 1$ and $d_s=1$.
However, the derivation of Eq.~\eqref{def_z_main} is not limited to these combinations. 
Here we fill the missing gap and, using the 2D and 3D Fibonacci models as examples, we show that there exist systems exhibiting critical dynamics with $d_l\neq 1$ and $ d_s\neq 1$.
We note that other examples of non-trivial critical dynamics in higher dimensional quasi-periodic models may also exist~\cite{Zhong98, Yuan00, Grimm02, Cerovski05, Thiem09, Thiem10, Lifshitz02, Sanchez04, Mandel08, Thiem09, Thiem10, Thiem12, Thiem13, Devakul17, Jagannathan21, Strkalj22} and should be studied in future work.

The separability of the potential in higher dimensions leads to the relation $\sigma^2_{\rm 1D}(t)=\sigma^2_{\rm 2D}(t)/2 = \sigma^2_{\rm 3D}(t)/3$ for the same values of $h$, see the End Matter, and thus to the identical dynamical exponents $z$ for the Fibonacci model in all dimensions.
This property is illustrated in Fig.$~$\ref{fig_example}, where we compare $\sigma^2(t)$ at $h=2$ in 2D and 1D, see Figs.~\ref{fig_example}(a) and~\ref{fig_example}(b), and at $h=4$ in 1D and 3D, see Figs.~\ref{fig_example}(c) and~\ref{fig_example}(d), observing identical $z$ at fixed $h$.
We hence extract $z$ from the results in the 1D Fibonacci model at the largest size $L=30000$.
Results for $1/z$ versus $h$ are shown in Fig.~\ref{fig_z}.
Starting with $z=1$ at $h=0$, the system transits from the superdiffusive regime, $1/2<1/z<1$, through the diffusive point at $1/z=1/2$, to the subdiffusive regime, $1/z<1/2$, in agreement with the results in Ref.~\cite{varma2019diffusive}.

In Fig.~\ref{fig_z}, we then compare the ratio $d_s/d_l$ to the inverse exponent $1/z$, and plot the results as a function of $h$.
In 1D, we observe $1/z \approx d_s/d_l$ for all values of $h$.
This suggests that Eq.~\eqref{def_z_main} is always valid and hence the entire parameter range of the 1D Fibonacci model exhibits critical dynamics~\cite{varma2019diffusive}.
In contrast, Eq.~\eqref{def_z_main} in the 2D and 3D Fibonacci models is only valid above a certain threshold potential $h^*$.
Determining $h^*$ as the lowest $h$ at which $1/z$ and $d_s/d_l$ match, we roughly estimate $h^*\approx 5$ in 2D and $h^* \approx 11.5$ in 3D.
At $h<h^*$, the dynamics lead to a completed delocalization of the wave-packet before the Heisenberg time is reached.

The emergence of a crossover scale $h=h^*$ for critical dynamics may appear puzzling provided that the eigenstates of separable Fibonacci models are critical (fractal) in any dimension for all $h$, see End Matter for details.
Our results hence raise a fundamental question: Are critical dynamics necessarily linked to fractal Hamiltonian eigenstates, or can one exist without the other?
We argue that if Eq.~\eqref{critical} defines critical transport, then fractal Hamiltonian eigenstates do not necessary require the dynamics to be critical. 
On the other hand, if one allows for critical dynamics to emerge beyond the validity of Eq.~\eqref{critical}, then the relation $z=d_l/d_s$ from Eq.~\eqref{def_z_main} may not be generally valid at criticality.
Here we propose the former interpretation of critical dynamics.
In this picture, validity of Eq.~\eqref{def_z_main} is the defining property of critical dynamics.

\textit{Conclusions.}---In this Letter, we introduced a novel perspective on critical dynamics in short-range quadratic Hamiltonians, grounded in the validity of the relation $z=d_l/d_s$.
We provided non-trivial examples in separable Fibonacci models, where both $d_s \neq 1$ and $d_l \neq 1$.
In this interpretation, at criticality, one does not observe superdiffusion in dimensions two or higher, and diffusion in dimensions three and higher.

\textit{Acknowlegment.}
We acknowledge discussions with P. Das, E. Ilievski, S. Jiricek, P. Prelovšek and T. Prosen.
We acknowledge support from the Slovenian Research and Innovation Agency (ARIS), Research core funding Grants No.~P1-0044, N1-0273 and J1-50005, as well as the Consolidator Grant Boundary-101126364 of the European Research Council (ERC).
We gratefully acknowledge the High Performance Computing Research Infrastructure Eastern Region (HCP RIVR) consortium~\cite{vega1} and European High Performance Computing Joint Undertaking (EuroHPC JU)~\cite{vega2}  for funding this research by providing computing resources of the
HPC system Vega at the Institute of Information sciences~\cite{vega3}.

\bibliographystyle{biblev1}
\bibliography{references,references1,references2}

\begin{thebibliography}{100}
\expandafter\ifx\csname url\endcsname\relax
  \def\url#1{{\tt #1}}\fi
\expandafter\ifx\csname urlprefix\endcsname\relax\def\urlprefix{URL }\fi
\expandafter\ifx\csname bibinfo\endcsname\relax\def\bibinfo#1#2{#2}\fi
\expandafter\ifx\csname eprint\endcsname\relax\def\eprint#1{\url{#1}}\fi

\bibitem{Datta_1995}
\bibinfo{author}{S.~Datta}, {\em \bibinfo{title}{Electronic Transport in
  Mesoscopic Systems}\/}, Cambridge Studies in Semiconductor Physics and
  Microelectronic Engineering (\bibinfo{publisher}{Cambridge University Press},
  \bibinfo{year}{1995}).

\bibitem{Ziman01}
\bibinfo{author}{J.~Ziman}, {\em \bibinfo{title}{Electrons and Phonons: The
  Theory of Transport Phenomena in Solids}\/} (\bibinfo{publisher}{Oxford
  University Press}, \bibinfo{year}{2001}).

\bibitem{Adams03}
\bibinfo{author}{D.~M. Adams}, \bibinfo{author}{L.~Brus},
  \bibinfo{author}{C.~E.~D. Chidsey}, \bibinfo{author}{S.~Creager},
  \bibinfo{author}{C.~Creutz}, \bibinfo{author}{C.~R. Kagan},
  \bibinfo{author}{P.~V. Kamat}, \bibinfo{author}{M.~Lieberman},
  \bibinfo{author}{S.~Lindsay}, \bibinfo{author}{R.~A. Marcus},
  \bibinfo{author}{R.~M. Metzger}, \bibinfo{author}{M.~E. Michel-Beyerle},
  \bibinfo{author}{J.~R. Miller}, \bibinfo{author}{M.~D. Newton},
  \bibinfo{author}{D.~R. Rolison}, \bibinfo{author}{O.~Sankey},
  \bibinfo{author}{K.~S. Schanze}, \bibinfo{author}{J.~Yardley}, and
  \bibinfo{author}{X.~Zhu}, \bibinfo{title}{Charge transfer on the nanoscale:
  Current status},
  \bibinfo{journal}{\href{http://dx.doi.org/10.1021/jp0268462}{J. Phys. Chem.
  B}} \href{http://dx.doi.org/10.1021/jp0268462}{{\bf \bibinfo{volume}{107}},
  \bibinfo{pages}{6668}}
  (\href{http://dx.doi.org/10.1021/jp0268462}{\bibinfo{year}{2003}}).

\bibitem{Oberhofer17}
\bibinfo{author}{H.~Oberhofer}, \bibinfo{author}{K.~Reuter}, and
  \bibinfo{author}{J.~Blumberger}, \bibinfo{title}{Charge transport in
  molecular materials: An assessment of computational methods},
  \bibinfo{journal}{\href{http://dx.doi.org/10.1021/acs.chemrev.7b00086}{Chem.
  Rev.}} \href{http://dx.doi.org/10.1021/acs.chemrev.7b00086}{{\bf
  \bibinfo{volume}{117}}, \bibinfo{pages}{10319}}
  (\href{http://dx.doi.org/10.1021/acs.chemrev.7b00086}{\bibinfo{year}{2017}}).

\bibitem{Mehrer07}
\bibinfo{author}{H.~Mehrer}, {\em \bibinfo{title}{Diffusion in Solids:
  Fundamentals, Methods, Materials, Diffusion-Controlled Processes}\/}
  (\bibinfo{publisher}{Springer Science \& Business Media},
  \bibinfo{year}{2007}).

\bibitem{Bokstein2018}
\bibinfo{author}{B.~S. Bokstein} and \bibinfo{author}{B.~B. Straumal}, {\em
  \bibinfo{title}{Diffusion in Materials Science and Technology}\/}, {\em
  \bibinfo{booktitle}{Diffusive Spreading in Nature, Technology and
  Society}\/}, edited by \bibinfo{editor}{A.~Bunde}, \bibinfo{editor}{J.~Caro},
  \bibinfo{editor}{J.~K{\"a}rger}, and \bibinfo{editor}{G.~Vogl},
  \bibinfo{pages}{261--275} (\bibinfo{publisher}{Springer International
  Publishing}, \bibinfo{address}{Cham}, \bibinfo{year}{2018}).

\bibitem{Bonitz98}
\bibinfo{author}{M.~Bonitz}, {\em \bibinfo{title}{Quantum Kinetic Theory}\/}
  (\bibinfo{publisher}{Springer}, \bibinfo{year}{1998}).

\bibitem{Thoss18}
\bibinfo{author}{M.~Thoss} and \bibinfo{author}{F.~Evers},
  \bibinfo{title}{Perspective: Theory of quantum transport in molecular
  junctions}, \bibinfo{journal}{\href{http://dx.doi.org/10.1063/1.5003306}{J.
  Chem. Phys.}} \href{http://dx.doi.org/10.1063/1.5003306}{{\bf
  \bibinfo{volume}{148}}, \bibinfo{pages}{030901}}
  (\href{http://dx.doi.org/10.1063/1.5003306}{\bibinfo{year}{2018}}).

\bibitem{Bertini21}
\bibinfo{author}{B.~Bertini}, \bibinfo{author}{F.~Heidrich-Meisner},
  \bibinfo{author}{C.~Karrasch}, \bibinfo{author}{T.~Prosen},
  \bibinfo{author}{R.~Steinigeweg}, and \bibinfo{author}{M.~\ifmmode
  \check{Z}\else \v{Z}\fi{}nidari\ifmmode~\check{c}\else \v{c}\fi{}},
  \bibinfo{title}{Finite-temperature transport in one-dimensional quantum
  lattice models},
  \bibinfo{journal}{\href{http://dx.doi.org/10.1103/RevModPhys.93.025003}{Rev.
  Mod. Phys.}} \href{http://dx.doi.org/10.1103/RevModPhys.93.025003}{{\bf
  \bibinfo{volume}{93}}, \bibinfo{pages}{025003}}
  (\href{http://dx.doi.org/10.1103/RevModPhys.93.025003}{\bibinfo{year}{2021}}).

\bibitem{waintal2024}
\bibinfo{author}{X.~Waintal}, \bibinfo{author}{M.~Wimmer},
  \bibinfo{author}{A.~Akhmerov}, \bibinfo{author}{C.~Groth},
  \bibinfo{author}{B.~K. Nikolic}, \bibinfo{author}{M.~Istas},
  \bibinfo{author}{T.~Örn Rosdahl}, and \bibinfo{author}{D.~Varjas},
  \bibinfo{title}{Computational quantum transport},
  \href{https://arxiv.org/abs/2407.16257}{\bibinfo{howpublished}{arXiv:2407.16257}}.

\bibitem{Steinigeweg14}
\bibinfo{author}{R.~Steinigeweg}, \bibinfo{author}{F.~Heidrich-Meisner},
  \bibinfo{author}{J.~Gemmer}, \bibinfo{author}{K.~Michielsen}, and
  \bibinfo{author}{H.~De~Raedt}, \bibinfo{title}{Scaling of diffusion constants
  in the spin-$\frac{1}{2}$ {XX} ladder},
  \bibinfo{journal}{\href{http://dx.doi.org/10.1103/PhysRevB.90.094417}{Phys.
  Rev. B}} \href{http://dx.doi.org/10.1103/PhysRevB.90.094417}{{\bf
  \bibinfo{volume}{90}}, \bibinfo{pages}{094417}}
  (\href{http://dx.doi.org/10.1103/PhysRevB.90.094417}{\bibinfo{year}{2014}}).

\bibitem{Karrasch14}
\bibinfo{author}{C.~Karrasch}, \bibinfo{author}{J.~E. Moore}, and
  \bibinfo{author}{F.~Heidrich-Meisner}, \bibinfo{title}{Real-time and
  real-space spin and energy dynamics in one-dimensional spin-$\frac{1}{2}$
  systems induced by local quantum quenches at finite temperatures},
  \bibinfo{journal}{\href{http://dx.doi.org/10.1103/PhysRevB.89.075139}{Phys.
  Rev. B}} \href{http://dx.doi.org/10.1103/PhysRevB.89.075139}{{\bf
  \bibinfo{volume}{89}}, \bibinfo{pages}{075139}}
  (\href{http://dx.doi.org/10.1103/PhysRevB.89.075139}{\bibinfo{year}{2014}}).

\bibitem{varma2019diffusive}
\bibinfo{author}{V.~K. Varma} and \bibinfo{author}{M.~\ifmmode \check{Z}\else
  \v{Z}\fi{}nidari\ifmmode~\check{c}\else \v{c}\fi{}},
  \bibinfo{title}{Diffusive transport in a quasiperiodic {F}ibonacci chain:
  Absence of many-body localization at weak interactions},
  \bibinfo{journal}{\href{http://dx.doi.org/10.1103/PhysRevB.100.085105}{Phys.
  Rev. B}} \href{http://dx.doi.org/10.1103/PhysRevB.100.085105}{{\bf
  \bibinfo{volume}{100}}, \bibinfo{pages}{085105}}
  (\href{http://dx.doi.org/10.1103/PhysRevB.100.085105}{\bibinfo{year}{2019}}).

\bibitem{Gopalakrishnan19}
\bibinfo{author}{S.~Gopalakrishnan} and \bibinfo{author}{R.~Vasseur},
  \bibinfo{title}{{Kinetic Theory of Spin Diffusion and Superdiffusion in
  {$XXZ$} Spin Chains}},
  \bibinfo{journal}{\href{http://dx.doi.org/10.1103/PhysRevLett.122.127202}{Phys.
  Rev. Lett.}} \href{http://dx.doi.org/10.1103/PhysRevLett.122.127202}{{\bf
  \bibinfo{volume}{122}}, \bibinfo{pages}{127202}}
  (\href{http://dx.doi.org/10.1103/PhysRevLett.122.127202}{\bibinfo{year}{2019}}).

\bibitem{DeNardis19}
\bibinfo{author}{J.~D. Nardis}, \bibinfo{author}{D.~Bernard}, and
  \bibinfo{author}{B.~Doyon}, \bibinfo{title}{{Diffusion in generalized
  hydrodynamics and quasiparticle scattering}},
  \bibinfo{journal}{\href{http://dx.doi.org/10.21468/SciPostPhys.6.4.049}{SciPost
  Phys.}} \href{http://dx.doi.org/10.21468/SciPostPhys.6.4.049}{{\bf
  \bibinfo{volume}{6}}, \bibinfo{pages}{049}}
  (\href{http://dx.doi.org/10.21468/SciPostPhys.6.4.049}{\bibinfo{year}{2019}}).

\bibitem{Richter19}
\bibinfo{author}{J.~Richter}, \bibinfo{author}{F.~Jin},
  \bibinfo{author}{L.~Knipschild}, \bibinfo{author}{J.~Herbrych},
  \bibinfo{author}{H.~De~Raedt}, \bibinfo{author}{K.~Michielsen},
  \bibinfo{author}{J.~Gemmer}, and \bibinfo{author}{R.~Steinigeweg},
  \bibinfo{title}{Magnetization and energy dynamics in spin ladders: Evidence
  of diffusion in time, frequency, position, and momentum},
  \bibinfo{journal}{\href{http://dx.doi.org/10.1103/PhysRevB.99.144422}{Phys.
  Rev. B}} \href{http://dx.doi.org/10.1103/PhysRevB.99.144422}{{\bf
  \bibinfo{volume}{99}}, \bibinfo{pages}{144422}}
  (\href{http://dx.doi.org/10.1103/PhysRevB.99.144422}{\bibinfo{year}{2019}}).

\bibitem{Wurtz20}
\bibinfo{author}{J.~Wurtz} and \bibinfo{author}{A.~Polkovnikov},
  \bibinfo{title}{Quantum diffusion in spin chains with phase space methods},
  \bibinfo{journal}{\href{http://dx.doi.org/10.1103/PhysRevE.101.052120}{Phys.
  Rev. E}} \href{http://dx.doi.org/10.1103/PhysRevE.101.052120}{{\bf
  \bibinfo{volume}{101}}, \bibinfo{pages}{052120}}
  (\href{http://dx.doi.org/10.1103/PhysRevE.101.052120}{\bibinfo{year}{2020}}).

\bibitem{Schubert21}
\bibinfo{author}{D.~Schubert}, \bibinfo{author}{J.~Richter},
  \bibinfo{author}{F.~Jin}, \bibinfo{author}{K.~Michielsen},
  \bibinfo{author}{H.~De~Raedt}, and \bibinfo{author}{R.~Steinigeweg},
  \bibinfo{title}{Quantum versus classical dynamics in spin models: Chains,
  ladders, and square lattices},
  \bibinfo{journal}{\href{http://dx.doi.org/10.1103/PhysRevB.104.054415}{Phys.
  Rev. B}} \href{http://dx.doi.org/10.1103/PhysRevB.104.054415}{{\bf
  \bibinfo{volume}{104}}, \bibinfo{pages}{054415}}
  (\href{http://dx.doi.org/10.1103/PhysRevB.104.054415}{\bibinfo{year}{2021}}).

\bibitem{Prelovsek21}
\bibinfo{author}{P.~Prelov\ifmmode~\check{s}\else \v{s}\fi{}ek} and
  \bibinfo{author}{J.~Herbrych}, \bibinfo{title}{Diffusion in the {A}nderson
  model in higher dimensions},
  \bibinfo{journal}{\href{http://dx.doi.org/10.1103/PhysRevB.103.L241107}{Phys.
  Rev. B}} \href{http://dx.doi.org/10.1103/PhysRevB.103.L241107}{{\bf
  \bibinfo{volume}{103}}, \bibinfo{pages}{L241107}}
  (\href{http://dx.doi.org/10.1103/PhysRevB.103.L241107}{\bibinfo{year}{2021}}).

\bibitem{Prelovsek22}
\bibinfo{author}{P.~Prelov\ifmmode~\check{s}\else \v{s}\fi{}ek},
  \bibinfo{author}{S.~Nandy}, \bibinfo{author}{Z.~Lenar\ifmmode \check{c}\else
  \v{c}\fi{}i\ifmmode~\check{c}\else \v{c}\fi{}},
  \bibinfo{author}{M.~Mierzejewski}, and \bibinfo{author}{J.~Herbrych},
  \bibinfo{title}{From dissipationless to normal diffusion in the easy-axis
  {H}eisenberg spin chain},
  \bibinfo{journal}{\href{http://dx.doi.org/10.1103/PhysRevB.106.245104}{Phys.
  Rev. B}} \href{http://dx.doi.org/10.1103/PhysRevB.106.245104}{{\bf
  \bibinfo{volume}{106}}, \bibinfo{pages}{245104}}
  (\href{http://dx.doi.org/10.1103/PhysRevB.106.245104}{\bibinfo{year}{2022}}).

\bibitem{Nandy23}
\bibinfo{author}{S.~Nandy}, \bibinfo{author}{Z.~Lenar\ifmmode \check{c}\else
  \v{c}\fi{}i\ifmmode~\check{c}\else \v{c}\fi{}},
  \bibinfo{author}{E.~Ilievski}, \bibinfo{author}{M.~Mierzejewski},
  \bibinfo{author}{J.~Herbrych}, and
  \bibinfo{author}{P.~Prelov\ifmmode~\check{s}\else \v{s}\fi{}ek},
  \bibinfo{title}{Spin diffusion in a perturbed isotropic {H}eisenberg spin
  chain},
  \bibinfo{journal}{\href{http://dx.doi.org/10.1103/PhysRevB.108.L081115}{Phys.
  Rev. B}} \href{http://dx.doi.org/10.1103/PhysRevB.108.L081115}{{\bf
  \bibinfo{volume}{108}}, \bibinfo{pages}{L081115}}
  (\href{http://dx.doi.org/10.1103/PhysRevB.108.L081115}{\bibinfo{year}{2023}}).

\bibitem{Prelovsek23b}
\bibinfo{author}{P.~Prelov\ifmmode~\check{s}\else \v{s}\fi{}ek},
  \bibinfo{author}{J.~Herbrych}, and \bibinfo{author}{M.~Mierzejewski},
  \bibinfo{title}{Slow diffusion and {T}houless localization criterion in
  modulated spin chains},
  \bibinfo{journal}{\href{http://dx.doi.org/10.1103/PhysRevB.108.035106}{Phys.
  Rev. B}} \href{http://dx.doi.org/10.1103/PhysRevB.108.035106}{{\bf
  \bibinfo{volume}{108}}, \bibinfo{pages}{035106}}
  (\href{http://dx.doi.org/10.1103/PhysRevB.108.035106}{\bibinfo{year}{2023}}).

\bibitem{Wang24}
\bibinfo{author}{J.~Wang}, \bibinfo{author}{M.~H. Lamann},
  \bibinfo{author}{R.~Steinigeweg}, and \bibinfo{author}{J.~Gemmer},
  \bibinfo{title}{Diffusion constants from the recursion method},
  \bibinfo{journal}{\href{http://dx.doi.org/10.1103/PhysRevB.110.104413}{Phys.
  Rev. B}} \href{http://dx.doi.org/10.1103/PhysRevB.110.104413}{{\bf
  \bibinfo{volume}{110}}, \bibinfo{pages}{104413}}
  (\href{http://dx.doi.org/10.1103/PhysRevB.110.104413}{\bibinfo{year}{2024}}).

\bibitem{kraft2024}
\bibinfo{author}{M.~Kraft}, \bibinfo{author}{M.~Kempa},
  \bibinfo{author}{J.~Wang}, \bibinfo{author}{S.~Nandy}, and
  \bibinfo{author}{R.~Steinigeweg}, \bibinfo{title}{Scaling of diffusion
  constants in perturbed easy-axis {H}eisenberg spin chains},
  \href{https://arxiv.org/abs/2410.22586}{\bibinfo{howpublished}{arXiv:2410.22586}}.

\bibitem{Ampelogiannis2025}
\bibinfo{author}{D.~Ampelogiannis} and \bibinfo{author}{B.~Doyon},
  \bibinfo{title}{Rigorous bound on hydrodynamic diffusion for chaotic open
  spin chains},
  \href{https://arxiv.org/abs/2501.07749}{\bibinfo{howpublished}{arXiv:2501.07749}}.

\bibitem{prosen24_short}
\bibinfo{author}{E.~Rosenberg}, \bibinfo{title}{{\it et al}, {Dynamics of
  magnetization at infinite temperature in a {H}eisenberg spin chain}},
  \bibinfo{journal}{\href{http://dx.doi.org/10.1126/science.adi7877}{Science}}
  \href{http://dx.doi.org/10.1126/science.adi7877}{{\bf \bibinfo{volume}{384}},
  \bibinfo{pages}{48}}
  (\href{http://dx.doi.org/10.1126/science.adi7877}{\bibinfo{year}{2024}}).

\bibitem{Znidaric11}
\bibinfo{author}{M.~\ifmmode \check{Z}\else
  \v{Z}\fi{}nidari\ifmmode~\check{c}\else \v{c}\fi{}}, \bibinfo{title}{{Spin
  Transport in a One-Dimensional Anisotropic {H}eisenberg Model}},
  \bibinfo{journal}{\href{http://dx.doi.org/10.1103/PhysRevLett.106.220601}{Phys.
  Rev. Lett.}} \href{http://dx.doi.org/10.1103/PhysRevLett.106.220601}{{\bf
  \bibinfo{volume}{106}}, \bibinfo{pages}{220601}}
  (\href{http://dx.doi.org/10.1103/PhysRevLett.106.220601}{\bibinfo{year}{2011}}).

\bibitem{Ljubotina17}
\bibinfo{author}{M.~Ljubotina}, \bibinfo{author}{M.~{\v Z}nidari{\v c}}, and
  \bibinfo{author}{T.~Prosen}, \bibinfo{title}{Spin diffusion from an
  inhomogeneous quench in an integrable system},
  \bibinfo{journal}{\href{http://dx.doi.org/10.1038/ncomms16117}{Nat. Commun.}}
  \href{http://dx.doi.org/10.1038/ncomms16117}{{\bf \bibinfo{volume}{8}},
  \bibinfo{pages}{16117}}
  (\href{http://dx.doi.org/10.1038/ncomms16117}{\bibinfo{year}{2017}}).

\bibitem{Ilievski18}
\bibinfo{author}{E.~Ilievski}, \bibinfo{author}{J.~De~Nardis},
  \bibinfo{author}{M.~Medenjak}, and \bibinfo{author}{T.~Prosen},
  \bibinfo{title}{{Superdiffusion in One-Dimensional Quantum Lattice Models}},
  \bibinfo{journal}{\href{http://dx.doi.org/10.1103/PhysRevLett.121.230602}{Phys.
  Rev. Lett.}} \href{http://dx.doi.org/10.1103/PhysRevLett.121.230602}{{\bf
  \bibinfo{volume}{121}}, \bibinfo{pages}{230602}}
  (\href{http://dx.doi.org/10.1103/PhysRevLett.121.230602}{\bibinfo{year}{2018}}).

\bibitem{Ljubotina19}
\bibinfo{author}{M.~Ljubotina}, \bibinfo{author}{M.~\ifmmode \check{Z}\else
  \v{Z}\fi{}nidari\ifmmode~\check{c}\else \v{c}\fi{}}, and
  \bibinfo{author}{T.~Prosen}, \bibinfo{title}{{{K}ardar-{P}arisi-{Z}hang
  Physics in the Quantum {H}eisenberg Magnet}},
  \bibinfo{journal}{\href{http://dx.doi.org/10.1103/PhysRevLett.122.210602}{Phys.
  Rev. Lett.}} \href{http://dx.doi.org/10.1103/PhysRevLett.122.210602}{{\bf
  \bibinfo{volume}{122}}, \bibinfo{pages}{210602}}
  (\href{http://dx.doi.org/10.1103/PhysRevLett.122.210602}{\bibinfo{year}{2019}}).

\bibitem{DeNardis20}
\bibinfo{author}{J.~De~Nardis}, \bibinfo{author}{S.~Gopalakrishnan},
  \bibinfo{author}{E.~Ilievski}, and \bibinfo{author}{R.~Vasseur},
  \bibinfo{title}{{Superdiffusion from Emergent Classical Solitons in Quantum
  Spin Chains}},
  \bibinfo{journal}{\href{http://dx.doi.org/10.1103/PhysRevLett.125.070601}{Phys.
  Rev. Lett.}} \href{http://dx.doi.org/10.1103/PhysRevLett.125.070601}{{\bf
  \bibinfo{volume}{125}}, \bibinfo{pages}{070601}}
  (\href{http://dx.doi.org/10.1103/PhysRevLett.125.070601}{\bibinfo{year}{2020}}).

\bibitem{Scheie21}
\bibinfo{author}{A.~Scheie}, \bibinfo{author}{N.~E. Sherman},
  \bibinfo{author}{M.~Dupont}, \bibinfo{author}{S.~E. Nagler},
  \bibinfo{author}{M.~B. Stone}, \bibinfo{author}{G.~E. Granroth},
  \bibinfo{author}{J.~E. Moore}, and \bibinfo{author}{D.~A. Tennant},
  \bibinfo{title}{Detection of {K}ardar--{P}arisi--{Z}hang hydrodynamics in a
  quantum {H}eisenberg spin-1/2 chain},
  \bibinfo{journal}{\href{http://dx.doi.org/https://doi.org/10.1038/s41567-021-01191-6}{Nat.
  Phys.}}
  \href{http://dx.doi.org/https://doi.org/10.1038/s41567-021-01191-6}{{\bf
  \bibinfo{volume}{17}}, \bibinfo{pages}{726}}
  (\href{http://dx.doi.org/https://doi.org/10.1038/s41567-021-01191-6}{\bibinfo{year}{2021}}).

\bibitem{Bulchandani21}
\bibinfo{author}{V.~B. Bulchandani}, \bibinfo{author}{S.~Gopalakrishnan}, and
  \bibinfo{author}{E.~Ilievski}, \bibinfo{title}{Superdiffusion in spin
  chains},
  \bibinfo{journal}{\href{http://dx.doi.org/10.1088/1742-5468/ac12c7}{J. Stat.
  Mech. Theor. Exp.}} \href{http://dx.doi.org/10.1088/1742-5468/ac12c7}{{\bf
  \bibinfo{volume}{2021}}, \bibinfo{pages}{084001}}
  (\href{http://dx.doi.org/10.1088/1742-5468/ac12c7}{\bibinfo{year}{2021}}).

\bibitem{Ilievski21}
\bibinfo{author}{E.~Ilievski}, \bibinfo{author}{J.~De~Nardis},
  \bibinfo{author}{S.~Gopalakrishnan}, \bibinfo{author}{R.~Vasseur}, and
  \bibinfo{author}{B.~Ware}, \bibinfo{title}{Superuniversality of
  superdiffusion},
  \bibinfo{journal}{\href{http://dx.doi.org/10.1103/PhysRevX.11.031023}{Phys.
  Rev. X}} \href{http://dx.doi.org/10.1103/PhysRevX.11.031023}{{\bf
  \bibinfo{volume}{11}}, \bibinfo{pages}{031023}}
  (\href{http://dx.doi.org/10.1103/PhysRevX.11.031023}{\bibinfo{year}{2021}}).

\bibitem{Wei22}
\bibinfo{author}{D.~Wei}, \bibinfo{author}{A.~Rubio-Abadal},
  \bibinfo{author}{B.~Ye}, \bibinfo{author}{F.~Machado},
  \bibinfo{author}{J.~Kemp}, \bibinfo{author}{K.~Srakaew},
  \bibinfo{author}{S.~Hollerith}, \bibinfo{author}{J.~Rui},
  \bibinfo{author}{S.~Gopalakrishnan}, \bibinfo{author}{N.~Y. Yao},
  \bibinfo{author}{I.~Bloch}, and \bibinfo{author}{J.~Zeiher},
  \bibinfo{title}{Quantum gas microscopy of {K}ardar-{P}arisi-{Z}hang
  superdiffusion},
  \bibinfo{journal}{\href{http://dx.doi.org/10.1126/science.abk2397}{Science}}
  \href{http://dx.doi.org/10.1126/science.abk2397}{{\bf \bibinfo{volume}{376}},
  \bibinfo{pages}{716}}
  (\href{http://dx.doi.org/10.1126/science.abk2397}{\bibinfo{year}{2022}}).

\bibitem{DeNardis23}
\bibinfo{author}{J.~De~Nardis}, \bibinfo{author}{S.~Gopalakrishnan}, and
  \bibinfo{author}{R.~Vasseur}, \bibinfo{title}{{Nonlinear Fluctuating
  Hydrodynamics for {K}ardar-{P}arisi-{Z}hang Scaling in Isotropic Spin
  Chains}},
  \bibinfo{journal}{\href{http://dx.doi.org/10.1103/PhysRevLett.131.197102}{Phys.
  Rev. Lett.}} \href{http://dx.doi.org/10.1103/PhysRevLett.131.197102}{{\bf
  \bibinfo{volume}{131}}, \bibinfo{pages}{197102}}
  (\href{http://dx.doi.org/10.1103/PhysRevLett.131.197102}{\bibinfo{year}{2023}}).

\bibitem{Krajnik24}
\bibinfo{author}{{\v Z}.~Krajnik}, \bibinfo{author}{J.~Schmidt},
  \bibinfo{author}{E.~Ilievski}, and \bibinfo{author}{T.~Prosen},
  \bibinfo{title}{{Dynamical Criticality of Magnetization Transfer in
  Integrable Spin Chains}},
  \bibinfo{journal}{\href{http://dx.doi.org/10.1103/PhysRevLett.132.017101}{Phys.
  Rev. Lett.}} \href{http://dx.doi.org/10.1103/PhysRevLett.132.017101}{{\bf
  \bibinfo{volume}{132}}, \bibinfo{pages}{017101}}
  (\href{http://dx.doi.org/10.1103/PhysRevLett.132.017101}{\bibinfo{year}{2024}}).

\bibitem{Gopalakrishnan24}
\bibinfo{author}{S.~Gopalakrishnan} and \bibinfo{author}{R.~Vasseur},
  \bibinfo{title}{Superdiffusion from nonabelian symmetries in nearly
  integrable systems},
  \bibinfo{journal}{\href{http://dx.doi.org/https://doi.org/10.1146/annurev-conmatphys-032922-110710}{Annu.
  Rev. Condens. Matter Phys.}}
  \href{http://dx.doi.org/https://doi.org/10.1146/annurev-conmatphys-032922-110710}{{\bf
  \bibinfo{volume}{15}}, \bibinfo{pages}{159}}
  (\href{http://dx.doi.org/https://doi.org/10.1146/annurev-conmatphys-032922-110710}{\bibinfo{year}{2024}}).

\bibitem{Bastianello24}
\bibinfo{author}{A.~Bastianello}, \bibinfo{author}{{\v Z}.~Krajnik}, and
  \bibinfo{author}{E.~Ilievski}, \bibinfo{title}{{Landau-{L}ifschitz Magnets:
  Exact Thermodynamics and Transport}},
  \bibinfo{journal}{\href{http://dx.doi.org/10.1103/PhysRevLett.133.107102}{Phys.
  Rev. Lett.}} \href{http://dx.doi.org/10.1103/PhysRevLett.133.107102}{{\bf
  \bibinfo{volume}{133}}, \bibinfo{pages}{107102}}
  (\href{http://dx.doi.org/10.1103/PhysRevLett.133.107102}{\bibinfo{year}{2024}}).

\bibitem{Kardar86}
\bibinfo{author}{M.~{K}ardar}, \bibinfo{author}{G.~{P}arisi}, and
  \bibinfo{author}{Y.-C. {Z}hang}, \bibinfo{title}{{Dynamic Scaling of Growing
  Interfaces}},
  \bibinfo{journal}{\href{http://dx.doi.org/10.1103/PhysRevLett.56.889}{Phys.
  Rev. Lett.}} \href{http://dx.doi.org/10.1103/PhysRevLett.56.889}{{\bf
  \bibinfo{volume}{56}}, \bibinfo{pages}{889}}
  (\href{http://dx.doi.org/10.1103/PhysRevLett.56.889}{\bibinfo{year}{1986}}).

\bibitem{Prelovsek17}
\bibinfo{author}{P.~Prelovšek}, \bibinfo{author}{M.~Mierzejewski},
  \bibinfo{author}{O.~Barišić}, and \bibinfo{author}{J.~Herbrych},
  \bibinfo{title}{Density correlations and transport in models of many-body
  localization},
  \bibinfo{journal}{\href{http://dx.doi.org/https://doi.org/10.1002/andp.201600362}{Ann.
  Phys.}} \href{http://dx.doi.org/https://doi.org/10.1002/andp.201600362}{{\bf
  \bibinfo{volume}{529}}, \bibinfo{pages}{1600362}}
  (\href{http://dx.doi.org/https://doi.org/10.1002/andp.201600362}{\bibinfo{year}{2017}}).

\bibitem{luitz_barlev_17}
\bibinfo{author}{D.~J. Luitz} and \bibinfo{author}{Y.~B. Lev},
  \bibinfo{title}{The ergodic side of the many-body localization transition},
  \bibinfo{journal}{\href{http://dx.doi.org/10.1002/andp.201600350}{Ann.
  Phys.}} \href{http://dx.doi.org/10.1002/andp.201600350}{{\bf
  \bibinfo{volume}{529}}, \bibinfo{pages}{1600350}}
  (\href{http://dx.doi.org/10.1002/andp.201600350}{\bibinfo{year}{2017}}).

\bibitem{Sierant_2025}
\bibinfo{author}{P.~Sierant}, \bibinfo{author}{M.~Lewenstein},
  \bibinfo{author}{A.~Scardicchio}, \bibinfo{author}{L.~Vidmar}, and
  \bibinfo{author}{J.~Zakrzewski}, \bibinfo{title}{Many-body localization in
  the age of classical computing},
  \bibinfo{journal}{\href{http://dx.doi.org/10.1088/1361-6633/ad9756}{Rep.
  Prog. Phys.}} \href{http://dx.doi.org/10.1088/1361-6633/ad9756}{{\bf
  \bibinfo{volume}{88}}, \bibinfo{pages}{026502}}
  (\href{http://dx.doi.org/10.1088/1361-6633/ad9756}{\bibinfo{year}{2025}}).

\bibitem{Chiaracane20}
\bibinfo{author}{C.~Chiaracane}, \bibinfo{author}{F.~Pietracaprina},
  \bibinfo{author}{A.~Purkayastha}, and \bibinfo{author}{J.~Goold},
  \bibinfo{title}{Quantum dynamics in the interacting {F}ibonacci chain},
  \bibinfo{journal}{\href{http://dx.doi.org/10.1103/PhysRevB.103.184205}{Phys.
  Rev. B}} \href{http://dx.doi.org/10.1103/PhysRevB.103.184205}{{\bf
  \bibinfo{volume}{103}}, \bibinfo{pages}{184205}}
  (\href{http://dx.doi.org/10.1103/PhysRevB.103.184205}{\bibinfo{year}{2021}}).

\bibitem{Caux_2011}
\bibinfo{author}{J.-S. Caux} and \bibinfo{author}{J.~Mossel},
  \bibinfo{title}{Remarks on the notion of quantum integrability},
  \bibinfo{journal}{\href{http://dx.doi.org/10.1088/1742-5468/2011/02/P02023}{J.
  Stat. Mech. Theor. Exp.}}
  \href{http://dx.doi.org/10.1088/1742-5468/2011/02/P02023}{{\bf
  \bibinfo{volume}{2011}}, \bibinfo{pages}{P02023}}
  (\href{http://dx.doi.org/10.1088/1742-5468/2011/02/P02023}{\bibinfo{year}{2011}}).

\bibitem{Calabrese_2016}
\bibinfo{author}{P.~Calabrese}, \bibinfo{author}{F.~H.~L. Essler}, and
  \bibinfo{author}{G.~Mussardo}, \bibinfo{title}{Introduction to ‘quantum
  integrability in out of equilibrium systems’},
  \bibinfo{journal}{\href{http://dx.doi.org/10.1088/1742-5468/2016/06/064001}{J.
  Stat. Mech. Theor. Exp.}}
  \href{http://dx.doi.org/10.1088/1742-5468/2016/06/064001}{{\bf
  \bibinfo{volume}{2016}}, \bibinfo{pages}{064001}}
  (\href{http://dx.doi.org/10.1088/1742-5468/2016/06/064001}{\bibinfo{year}{2016}}).

\bibitem{anderson_58}
\bibinfo{author}{P.~W. {A}nderson}, \bibinfo{title}{Absence of diffusion in
  certain random lattices},
  \bibinfo{journal}{\href{http://dx.doi.org/10.1103/PhysRev.109.1492}{Phys.
  Rev.}} \href{http://dx.doi.org/10.1103/PhysRev.109.1492}{{\bf
  \bibinfo{volume}{109}}, \bibinfo{pages}{1492}}
  (\href{http://dx.doi.org/10.1103/PhysRev.109.1492}{\bibinfo{year}{1958}}).

\bibitem{Frenkel02}
{\em \bibinfo{booktitle}{Understanding Molecular Simulation (Second
  Edition)}\/}, edited by \bibinfo{editor}{D.~Frenkel} and
  \bibinfo{editor}{B.~Smit} (\bibinfo{publisher}{Academic Press},
  \bibinfo{address}{San Diego}, \bibinfo{year}{2002}).

\bibitem{Troisi06}
\bibinfo{author}{A.~Troisi} and \bibinfo{author}{G.~Orlandi},
  \bibinfo{title}{{Charge-Transport Regime of Crystalline Organic
  Semiconductors: Diffusion Limited by Thermal Off-Diagonal Electronic
  Disorder}},
  \bibinfo{journal}{\href{http://dx.doi.org/10.1103/PhysRevLett.96.086601}{Phys.
  Rev. Lett.}} \href{http://dx.doi.org/10.1103/PhysRevLett.96.086601}{{\bf
  \bibinfo{volume}{96}}, \bibinfo{pages}{086601}}
  (\href{http://dx.doi.org/10.1103/PhysRevLett.96.086601}{\bibinfo{year}{2006}}).

\bibitem{Troisi11}
\bibinfo{author}{A.~Troisi}, \bibinfo{title}{{Dynamic disorder in molecular
  semiconductors: Charge transport in two dimensions}},
  \bibinfo{journal}{\href{http://dx.doi.org/10.1063/1.3524314}{J. Chem. Phys}}
  \href{http://dx.doi.org/10.1063/1.3524314}{{\bf \bibinfo{volume}{134}},
  \bibinfo{pages}{034702}}
  (\href{http://dx.doi.org/10.1063/1.3524314}{\bibinfo{year}{2011}}).

\bibitem{Wang11}
\bibinfo{author}{L.~Wang}, \bibinfo{author}{D.~Beljonne},
  \bibinfo{author}{L.~Chen}, and \bibinfo{author}{Q.~Shi},
  \bibinfo{title}{Mixed quantum-classical simulations of charge transport in
  organic materials: Numerical benchmark of the {S}u-{S}chrieffer-{H}eeger
  model}, \bibinfo{journal}{\href{http://dx.doi.org/10.1063/1.3604561}{J. Chem.
  Phys}} \href{http://dx.doi.org/10.1063/1.3604561}{{\bf
  \bibinfo{volume}{134}}, \bibinfo{pages}{244116}}
  (\href{http://dx.doi.org/10.1063/1.3604561}{\bibinfo{year}{2011}}).

\bibitem{Chen17a}
\bibinfo{author}{L.~Chen} and \bibinfo{author}{Y.~Zhao}, \bibinfo{title}{Finite
  temperature dynamics of a {H}olstein polaron: The thermo-field dynamics
  approach}, \bibinfo{journal}{\href{http://dx.doi.org/10.1063/1.5000823}{J.
  Chem. Phys}} \href{http://dx.doi.org/10.1063/1.5000823}{{\bf
  \bibinfo{volume}{147}}, \bibinfo{pages}{214102}}
  (\href{http://dx.doi.org/10.1063/1.5000823}{\bibinfo{year}{2017}}).

\bibitem{Kloss19}
\bibinfo{author}{B.~Kloss}, \bibinfo{author}{D.~R. Reichman}, and
  \bibinfo{author}{R.~Tempelaar}, \bibinfo{title}{{Multiset Matrix Product
  State Calculations Reveal Mobile Franck-Condon Excitations Under Strong
  {H}olstein-Type Coupling}},
  \bibinfo{journal}{\href{http://dx.doi.org/10.1103/PhysRevLett.123.126601}{Phys.
  Rev. Lett.}} \href{http://dx.doi.org/10.1103/PhysRevLett.123.126601}{{\bf
  \bibinfo{volume}{123}}, \bibinfo{pages}{126601}}
  (\href{http://dx.doi.org/10.1103/PhysRevLett.123.126601}{\bibinfo{year}{2019}}).

\bibitem{tenBrink22}
\bibinfo{author}{M.~ten Brink}, \bibinfo{author}{S.~Gräber},
  \bibinfo{author}{M.~Hopjan}, \bibinfo{author}{D.~Jansen},
  \bibinfo{author}{J.~Stolpp}, \bibinfo{author}{F.~Heidrich-Meisner}, and
  \bibinfo{author}{P.~E. Blöchl}, \bibinfo{title}{Real-time non-adiabatic
  dynamics in the one-dimensional {H}olstein model: Trajectory-based vs exact
  methods}, \bibinfo{journal}{\href{http://dx.doi.org/10.1063/5.0092063}{J.
  Chem. Phys}} \href{http://dx.doi.org/10.1063/5.0092063}{{\bf
  \bibinfo{volume}{156}}, \bibinfo{pages}{234109}}
  (\href{http://dx.doi.org/10.1063/5.0092063}{\bibinfo{year}{2022}}).

\bibitem{Khan21}
\bibinfo{author}{M.~M. Khan}, \bibinfo{author}{H.~Ter\ifmmode~\mbox{\c{c}}\else
  \c{c}\fi{}as}, \bibinfo{author}{J.~T. Mendon\ifmmode~\mbox{\c{c}}\else
  \c{c}\fi{}a}, \bibinfo{author}{J.~Wehr}, \bibinfo{author}{C.~Charalambous},
  \bibinfo{author}{M.~Lewenstein}, and \bibinfo{author}{M.~A. Garcia-March},
  \bibinfo{title}{Quantum dynamics of a {B}ose polaron in a $d$-dimensional
  {B}ose-{E}instein condensate},
  \bibinfo{journal}{\href{http://dx.doi.org/10.1103/PhysRevA.103.023303}{Phys.
  Rev. A}} \href{http://dx.doi.org/10.1103/PhysRevA.103.023303}{{\bf
  \bibinfo{volume}{103}}, \bibinfo{pages}{023303}}
  (\href{http://dx.doi.org/10.1103/PhysRevA.103.023303}{\bibinfo{year}{2021}}).

\bibitem{Birschitzky24}
\bibinfo{author}{V.~C. Birschitzky}, \bibinfo{author}{L.~Leoni},
  \bibinfo{author}{M.~Reticcioli}, and \bibinfo{author}{C.~Franchini},
  \bibinfo{title}{Machine learning small polaron dynamics},
  \href{https://arxiv.org/abs/2409.16179}{\bibinfo{howpublished}{arXiv:2409.16179}}.

\bibitem{Marquardt21}
\bibinfo{author}{R.~Marquardt}, \bibinfo{title}{Mean square displacement of a
  free quantum particle in a thermal state},
  \bibinfo{journal}{\href{http://dx.doi.org/10.1080/00268976.2024.2322023}{Mol.
  Phys.}} \href{http://dx.doi.org/10.1080/00268976.2024.2322023}{{\bf
  \bibinfo{volume}{119}}, \bibinfo{pages}{e1971315}}
  (\href{http://dx.doi.org/10.1080/00268976.2024.2322023}{\bibinfo{year}{2021}}).

\bibitem{Schirripa24}
\bibinfo{author}{C.~Schirripa~Spagnolo} and \bibinfo{author}{S.~Luin},
  \bibinfo{title}{Trajectory analysis in single-particle tracking: From mean
  squared displacement to machine learning approaches},
  \bibinfo{journal}{\href{http://dx.doi.org/10.3390/ijms25168660}{Int. J. Mol.
  Sci.}} \href{http://dx.doi.org/10.3390/ijms25168660}{{\bf
  \bibinfo{volume}{25}}}
  (\href{http://dx.doi.org/10.3390/ijms25168660}{\bibinfo{year}{2024}}).

\bibitem{Bindech24}
\bibinfo{author}{O.~Bindech}, \bibinfo{author}{F.~Gatti},
  \bibinfo{author}{S.~Mandal}, \bibinfo{author}{R.~Marquardt},
  \bibinfo{author}{S.~L.}, and \bibinfo{author}{J.~C. Tremblay},
  \bibinfo{title}{The mean square displacement of a ballistic quantum
  particle},
  \bibinfo{journal}{\href{http://dx.doi.org/10.1080/00268976.2024.2322023}{Mol.
  Phys.}} \href{http://dx.doi.org/10.1080/00268976.2024.2322023}{{\bf
  \bibinfo{volume}{122}}, \bibinfo{pages}{e2322023}}
  (\href{http://dx.doi.org/10.1080/00268976.2024.2322023}{\bibinfo{year}{2024}}).

\bibitem{Fujimoto21}
\bibinfo{author}{K.~Fujimoto}, \bibinfo{author}{R.~Hamazaki}, and
  \bibinfo{author}{Y.~Kawaguchi}, \bibinfo{title}{{Dynamical Scaling of Surface
  Roughness and Entanglement Entropy in Disordered Fermion Models}},
  \bibinfo{journal}{\href{http://dx.doi.org/10.1103/PhysRevLett.127.090601}{Phys.
  Rev. Lett.}} \href{http://dx.doi.org/10.1103/PhysRevLett.127.090601}{{\bf
  \bibinfo{volume}{127}}, \bibinfo{pages}{090601}}
  (\href{http://dx.doi.org/10.1103/PhysRevLett.127.090601}{\bibinfo{year}{2021}}).

\bibitem{Bhakuni24}
\bibinfo{author}{D.~S. Bhakuni} and \bibinfo{author}{Y.~B. Lev},
  \bibinfo{title}{Dynamic scaling relation in quantum many-body systems},
  \bibinfo{journal}{\href{http://dx.doi.org/10.1103/PhysRevB.110.014203}{Phys.
  Rev. B}} \href{http://dx.doi.org/10.1103/PhysRevB.110.014203}{{\bf
  \bibinfo{volume}{110}}, \bibinfo{pages}{014203}}
  (\href{http://dx.doi.org/10.1103/PhysRevB.110.014203}{\bibinfo{year}{2024}}).

\bibitem{Sreemayee24}
\bibinfo{author}{S.~Aditya} and \bibinfo{author}{N.~Roy},
  \bibinfo{title}{Family-{V}icsek dynamical scaling and
  {K}ardar-{P}arisi-{Z}hang-like superdiffusive growth of surface roughness in
  a driven one-dimensional quasiperiodic model},
  \bibinfo{journal}{\href{http://dx.doi.org/10.1103/PhysRevB.109.035164}{Phys.
  Rev. B}} \href{http://dx.doi.org/10.1103/PhysRevB.109.035164}{{\bf
  \bibinfo{volume}{109}}, \bibinfo{pages}{035164}}
  (\href{http://dx.doi.org/10.1103/PhysRevB.109.035164}{\bibinfo{year}{2024}}).

\bibitem{Ketzmerick92}
\bibinfo{author}{R.~Ketzmerick}, \bibinfo{author}{G.~Petschel}, and
  \bibinfo{author}{T.~Geisel}, \bibinfo{title}{{Slow decay of temporal
  correlations in quantum systems with {C}antor spectra}},
  \bibinfo{journal}{\href{http://dx.doi.org/10.1103/PhysRevLett.69.695}{Phys.
  Rev. Lett.}} \href{http://dx.doi.org/10.1103/PhysRevLett.69.695}{{\bf
  \bibinfo{volume}{69}}, \bibinfo{pages}{695}}
  (\href{http://dx.doi.org/10.1103/PhysRevLett.69.695}{\bibinfo{year}{1992}}).

\bibitem{Ketzmerick97}
\bibinfo{author}{R.~Ketzmerick}, \bibinfo{author}{K.~Kruse},
  \bibinfo{author}{S.~Kraut}, and \bibinfo{author}{T.~Geisel},
  \bibinfo{title}{{What Determines the Spreading of a Wave Packet?}},
  \bibinfo{journal}{\href{http://dx.doi.org/10.1103/PhysRevLett.79.1959}{Phys.
  Rev. Lett.}} \href{http://dx.doi.org/10.1103/PhysRevLett.79.1959}{{\bf
  \bibinfo{volume}{79}}, \bibinfo{pages}{1959}}
  (\href{http://dx.doi.org/10.1103/PhysRevLett.79.1959}{\bibinfo{year}{1997}}).

\bibitem{Thiem13a}
\bibinfo{author}{S.~Thiem} and \bibinfo{author}{M.~Schreiber},
  \bibinfo{title}{Quantum diffusion in separable d-dimensional quasiperiodic
  tilings}, {\em \bibinfo{booktitle}{Aperiodic Crystals}\/}, edited by
  \bibinfo{editor}{S.~Schmid}, \bibinfo{editor}{R.~L. Withers}, and
  \bibinfo{editor}{R.~Lifshitz}, \bibinfo{pages}{89--94}
  (\bibinfo{publisher}{Springer Netherlands}, \bibinfo{address}{Dordrecht},
  \bibinfo{year}{2013}).

\bibitem{hopjan2023}
\bibinfo{author}{M.~Hopjan} and \bibinfo{author}{L.~Vidmar},
  \bibinfo{title}{{Scale-Invariant Survival Probability at Eigenstate
  Transitions}},
  \bibinfo{journal}{\href{http://dx.doi.org/10.1103/PhysRevLett.131.060404}{Phys.
  Rev. Lett.}} \href{http://dx.doi.org/10.1103/PhysRevLett.131.060404}{{\bf
  \bibinfo{volume}{131}}, \bibinfo{pages}{060404}}
  (\href{http://dx.doi.org/10.1103/PhysRevLett.131.060404}{\bibinfo{year}{2023}}).

\bibitem{Kohmoto83}
\bibinfo{author}{M.~Kohmoto}, \bibinfo{title}{{Metal-Insulator Transition and
  Scaling for Incommensurate Systems}},
  \bibinfo{journal}{\href{http://dx.doi.org/10.1103/PhysRevLett.51.1198}{Phys.
  Rev. Lett.}} \href{http://dx.doi.org/10.1103/PhysRevLett.51.1198}{{\bf
  \bibinfo{volume}{51}}, \bibinfo{pages}{1198}}
  (\href{http://dx.doi.org/10.1103/PhysRevLett.51.1198}{\bibinfo{year}{1983}}).

\bibitem{Kohmoto84}
\bibinfo{author}{M.~Kohmoto} and \bibinfo{author}{Y.~Oono},
  \bibinfo{title}{{C}antor spectrum for an almost periodic {S}chrödinger
  equation and a dynamical map},
  \bibinfo{journal}{\href{http://dx.doi.org/https://doi.org/10.1016/0375-9601(84)90928-9}{Phys.
  Lett. A}}
  \href{http://dx.doi.org/https://doi.org/10.1016/0375-9601(84)90928-9}{{\bf
  \bibinfo{volume}{102}}, \bibinfo{pages}{145}}
  (\href{http://dx.doi.org/https://doi.org/10.1016/0375-9601(84)90928-9}{\bibinfo{year}{1984}}).

\bibitem{Tang86}
\bibinfo{author}{C.~Tang} and \bibinfo{author}{M.~Kohmoto},
  \bibinfo{title}{Global scaling properties of the spectrum for a quasiperiodic
  {s}chr\"odinger equation},
  \bibinfo{journal}{\href{http://dx.doi.org/10.1103/PhysRevB.34.2041}{Phys.
  Rev. B}} \href{http://dx.doi.org/10.1103/PhysRevB.34.2041}{{\bf
  \bibinfo{volume}{34}}, \bibinfo{pages}{2041}}
  (\href{http://dx.doi.org/10.1103/PhysRevB.34.2041}{\bibinfo{year}{1986}}).

\bibitem{Halsey86}
\bibinfo{author}{T.~C. Halsey}, \bibinfo{author}{M.~H. Jensen},
  \bibinfo{author}{L.~P. Kadanoff}, \bibinfo{author}{I.~Procaccia}, and
  \bibinfo{author}{B.~I. Shraiman}, \bibinfo{title}{Fractal measures and their
  singularities: The characterization of strange sets},
  \bibinfo{journal}{\href{http://dx.doi.org/10.1103/PhysRevA.33.1141}{Phys.
  Rev. A}} \href{http://dx.doi.org/10.1103/PhysRevA.33.1141}{{\bf
  \bibinfo{volume}{33}}, \bibinfo{pages}{1141}}
  (\href{http://dx.doi.org/10.1103/PhysRevA.33.1141}{\bibinfo{year}{1986}}).

\bibitem{Kohmoto87}
\bibinfo{author}{M.~Kohmoto}, \bibinfo{author}{B.~Sutherland}, and
  \bibinfo{author}{C.~Tang}, \bibinfo{title}{Critical wave functions and a
  {C}antor-set spectrum of a one-dimensional quasicrystal model},
  \bibinfo{journal}{\href{http://dx.doi.org/10.1103/PhysRevB.35.1020}{Phys.
  Rev. B}} \href{http://dx.doi.org/10.1103/PhysRevB.35.1020}{{\bf
  \bibinfo{volume}{35}}, \bibinfo{pages}{1020}}
  (\href{http://dx.doi.org/10.1103/PhysRevB.35.1020}{\bibinfo{year}{1987}}).

\bibitem{Abe87}
\bibinfo{author}{S.~Abe} and \bibinfo{author}{H.~Hiramoto},
  \bibinfo{title}{Fractal dynamics of electron wave packets in one-dimensional
  quasiperiodic systems},
  \bibinfo{journal}{\href{http://dx.doi.org/10.1103/PhysRevA.36.5349}{Phys.
  Rev. A}} \href{http://dx.doi.org/10.1103/PhysRevA.36.5349}{{\bf
  \bibinfo{volume}{36}}, \bibinfo{pages}{5349}}
  (\href{http://dx.doi.org/10.1103/PhysRevA.36.5349}{\bibinfo{year}{1987}}).

\bibitem{Hiramoto88a}
\bibinfo{author}{H.~Hiramoto} and \bibinfo{author}{S.~Abe},
  \bibinfo{title}{Dynamics of an electron in quasiperiodic systems. i.
  {F}ibonacci model},
  \bibinfo{journal}{\href{http://dx.doi.org/10.1143/JPSJ.57.230}{J. Phys. Soc.
  Jpn.}} \href{http://dx.doi.org/10.1143/JPSJ.57.230}{{\bf
  \bibinfo{volume}{57}}, \bibinfo{pages}{230}}
  (\href{http://dx.doi.org/10.1143/JPSJ.57.230}{\bibinfo{year}{1988}}).

\bibitem{Hiramoto88b}
\bibinfo{author}{H.~Hiramoto} and \bibinfo{author}{S.~Abe},
  \bibinfo{title}{Dynamics of an electron in quasiperiodic systems. ii.
  {H}arper's model},
  \bibinfo{journal}{\href{http://dx.doi.org/10.1143/JPSJ.57.1365}{J. Phys. Soc.
  Jpn.}} \href{http://dx.doi.org/10.1143/JPSJ.57.1365}{{\bf
  \bibinfo{volume}{57}}, \bibinfo{pages}{1365}}
  (\href{http://dx.doi.org/10.1143/JPSJ.57.1365}{\bibinfo{year}{1988}}).

\bibitem{Guarneri89}
\bibinfo{author}{I.~Guarneri}, \bibinfo{title}{{Spectral Properties of Quantum
  Diffusion on Discrete Lattices}},
  \bibinfo{journal}{\href{http://dx.doi.org/10.1209/0295-5075/10/2/001}{Europhys.
  Lett.}} \href{http://dx.doi.org/10.1209/0295-5075/10/2/001}{{\bf
  \bibinfo{volume}{10}}, \bibinfo{pages}{95}}
  (\href{http://dx.doi.org/10.1209/0295-5075/10/2/001}{\bibinfo{year}{1989}}).

\bibitem{Geisel91}
\bibinfo{author}{T.~Geisel}, \bibinfo{author}{R.~Ketzmerick}, and
  \bibinfo{author}{G.~Petschel}, \bibinfo{title}{{New class of level statistics
  in quantum systems with unbounded diffusion}},
  \bibinfo{journal}{\href{http://dx.doi.org/10.1103/PhysRevLett.66.1651}{Phys.
  Rev. Lett.}} \href{http://dx.doi.org/10.1103/PhysRevLett.66.1651}{{\bf
  \bibinfo{volume}{66}}, \bibinfo{pages}{1651}}
  (\href{http://dx.doi.org/10.1103/PhysRevLett.66.1651}{\bibinfo{year}{1991}}).

\bibitem{Geisel91b}
\bibinfo{author}{T.~Geisel}, \bibinfo{author}{R.~Ketzmerick}, and
  \bibinfo{author}{G.~Petschel}, \bibinfo{title}{Metamorphosis of a {C}antor
  spectrum due to classical chaos},
  \bibinfo{journal}{\href{http://dx.doi.org/10.1103/PhysRevLett.67.3635}{Phys.
  Rev. Lett.}} \href{http://dx.doi.org/10.1103/PhysRevLett.67.3635}{{\bf
  \bibinfo{volume}{67}}, \bibinfo{pages}{3635}}
  (\href{http://dx.doi.org/10.1103/PhysRevLett.67.3635}{\bibinfo{year}{1991}}).

\bibitem{Geisel1992}
\bibinfo{author}{T.~Geisel}, \bibinfo{author}{R.~Ketzmerick}, and
  \bibinfo{author}{G.~Petschel}, {\em \bibinfo{title}{Unbounded Quantum
  Diffusion and a New Class of Level Statistics}\/}, {\em
  \bibinfo{booktitle}{Quantum Chaos --- Quantum Measurement}\/}, edited by
  \bibinfo{editor}{P.~Cvitanovi{\'{c}}}, \bibinfo{editor}{I.~Percival}, and
  \bibinfo{editor}{A.~Wirzba}, \bibinfo{pages}{43--59}
  (\bibinfo{publisher}{Springer Netherlands}, \bibinfo{address}{Dordrecht},
  \bibinfo{year}{1992}).

\bibitem{Lima91}
\bibinfo{author}{R.~Lima} and \bibinfo{author}{D.~Shepelyansky},
  \bibinfo{title}{Fast delocalization in a model of quantum kicked rotator},
  \bibinfo{journal}{\href{http://dx.doi.org/10.1103/PhysRevLett.67.1377}{Phys.
  Rev. Lett.}} \href{http://dx.doi.org/10.1103/PhysRevLett.67.1377}{{\bf
  \bibinfo{volume}{67}}, \bibinfo{pages}{1377}}
  (\href{http://dx.doi.org/10.1103/PhysRevLett.67.1377}{\bibinfo{year}{1991}}).

\bibitem{Artuso92a}
\bibinfo{author}{R.~Artuso}, \bibinfo{author}{G.~Casati}, and
  \bibinfo{author}{D.~Shepelyansky}, \bibinfo{title}{Fractal spectrum and
  anomalous diffusion in the kicked {H}arper model},
  \bibinfo{journal}{\href{http://dx.doi.org/10.1103/PhysRevLett.68.3826}{Phys.
  Rev. Lett.}} \href{http://dx.doi.org/10.1103/PhysRevLett.68.3826}{{\bf
  \bibinfo{volume}{68}}, \bibinfo{pages}{3826}}
  (\href{http://dx.doi.org/10.1103/PhysRevLett.68.3826}{\bibinfo{year}{1992}}).

\bibitem{Artuso92b}
\bibinfo{author}{R.~Artuso}, \bibinfo{author}{F.~Borgonovi},
  \bibinfo{author}{I.~Guarneri}, \bibinfo{author}{L.~Rebuzzini}, and
  \bibinfo{author}{G.~Casati}, \bibinfo{title}{Phase diagram in the kicked
  {H}arper model},
  \bibinfo{journal}{\href{http://dx.doi.org/10.1103/PhysRevLett.69.3302}{Phys.
  Rev. Lett.}} \href{http://dx.doi.org/10.1103/PhysRevLett.69.3302}{{\bf
  \bibinfo{volume}{69}}, \bibinfo{pages}{3302}}
  (\href{http://dx.doi.org/10.1103/PhysRevLett.69.3302}{\bibinfo{year}{1992}}).

\bibitem{Guarneri93}
\bibinfo{author}{I.~Guarneri}, \bibinfo{title}{{On an Estimate Concerning
  Quantum Diffusion in the Presence of a Fractal Spectrum}},
  \bibinfo{journal}{\href{http://dx.doi.org/10.1209/0295-5075/21/7/003}{Europhys.
  Lett.}} \href{http://dx.doi.org/10.1209/0295-5075/21/7/003}{{\bf
  \bibinfo{volume}{21}}, \bibinfo{pages}{729}}
  (\href{http://dx.doi.org/10.1209/0295-5075/21/7/003}{\bibinfo{year}{1993}}).

\bibitem{Evangelou93}
\bibinfo{author}{S.~N. Evangelou} and \bibinfo{author}{D.~E. Katsanos},
  \bibinfo{title}{Multifractal quantum evolution at a mobility edge},
  \bibinfo{journal}{\href{http://dx.doi.org/10.1088/0305-4470/26/23/010}{J.
  Phys. A Math. Gen.}}
  \href{http://dx.doi.org/10.1088/0305-4470/26/23/010}{{\bf
  \bibinfo{volume}{26}}, \bibinfo{pages}{L1243}}
  (\href{http://dx.doi.org/10.1088/0305-4470/26/23/010}{\bibinfo{year}{1993}}).

\bibitem{Guarneri94}
\bibinfo{author}{I.~Guarneri} and \bibinfo{author}{G.~Mantica},
  \bibinfo{title}{{Multifractal Energy Spectra and Their Dynamical
  Implications}},
  \bibinfo{journal}{\href{http://dx.doi.org/10.1103/PhysRevLett.73.3379}{Phys.
  Rev. Lett.}} \href{http://dx.doi.org/10.1103/PhysRevLett.73.3379}{{\bf
  \bibinfo{volume}{73}}, \bibinfo{pages}{3379}}
  (\href{http://dx.doi.org/10.1103/PhysRevLett.73.3379}{\bibinfo{year}{1994}}).

\bibitem{Wilkinson94}
\bibinfo{author}{M.~Wilkinson} and \bibinfo{author}{E.~J. Austin},
  \bibinfo{title}{Spectral dimension and dynamics for {H}arper's equation},
  \bibinfo{journal}{\href{http://dx.doi.org/10.1103/PhysRevB.50.1420}{Phys.
  Rev. B}} \href{http://dx.doi.org/10.1103/PhysRevB.50.1420}{{\bf
  \bibinfo{volume}{50}}, \bibinfo{pages}{1420}}
  (\href{http://dx.doi.org/10.1103/PhysRevB.50.1420}{\bibinfo{year}{1994}}).

\bibitem{Fleischmann95}
\bibinfo{author}{R.~Fleischmann}, \bibinfo{author}{T.~Geisel},
  \bibinfo{author}{R.~Ketzmerick}, and \bibinfo{author}{G.~Petschel},
  \bibinfo{title}{Quantum diffusion, fractal spectra, and chaos in
  semiconductor microstructures},
  \bibinfo{journal}{\href{http://dx.doi.org/https://doi.org/10.1016/0167-2789(95)00098-O}{Phys.
  D Nonlinear Phenom.}}
  \href{http://dx.doi.org/https://doi.org/10.1016/0167-2789(95)00098-O}{{\bf
  \bibinfo{volume}{86}}, \bibinfo{pages}{171}}
  (\href{http://dx.doi.org/https://doi.org/10.1016/0167-2789(95)00098-O}{\bibinfo{year}{1995}}).

\bibitem{Guarneri95}
\bibinfo{author}{I.~Guarneri} and \bibinfo{author}{M.~D. Meo},
  \bibinfo{title}{Fractal spectrum of a quasi-periodically driven spin system},
  \bibinfo{journal}{\href{http://dx.doi.org/10.1088/0305-4470/28/10/005}{J.
  Phys. A Math. Gen.}}
  \href{http://dx.doi.org/10.1088/0305-4470/28/10/005}{{\bf
  \bibinfo{volume}{28}}, \bibinfo{pages}{2717}}
  (\href{http://dx.doi.org/10.1088/0305-4470/28/10/005}{\bibinfo{year}{1995}}).

\bibitem{Zhong95}
\bibinfo{author}{J.~X. Zhong} and \bibinfo{author}{R.~Mosseri},
  \bibinfo{title}{Quantum dynamics in quasiperiodic systems},
  \bibinfo{journal}{\href{http://dx.doi.org/10.1088/0953-8984/7/44/008}{J.
  Condens. Matter Phys.}}
  \href{http://dx.doi.org/10.1088/0953-8984/7/44/008}{{\bf
  \bibinfo{volume}{7}}, \bibinfo{pages}{8383}}
  (\href{http://dx.doi.org/10.1088/0953-8984/7/44/008}{\bibinfo{year}{1995}}).

\bibitem{Kawarabayashi95}
\bibinfo{author}{T.~Kawarabayashi} and \bibinfo{author}{T.~Ohtsuki},
  \bibinfo{title}{Diffusion of electrons in random magnetic fields},
  \bibinfo{journal}{\href{http://dx.doi.org/10.1103/PhysRevB.51.10897}{Phys.
  Rev. B}} \href{http://dx.doi.org/10.1103/PhysRevB.51.10897}{{\bf
  \bibinfo{volume}{51}}, \bibinfo{pages}{10897}}
  (\href{http://dx.doi.org/10.1103/PhysRevB.51.10897}{\bibinfo{year}{1995}}).

\bibitem{Picheon96}
\bibinfo{author}{F.~Pi\'echon}, \bibinfo{title}{{Anomalous Diffusion Properties
  of Wave Packets on Quasiperiodic Chains}},
  \bibinfo{journal}{\href{http://dx.doi.org/10.1103/PhysRevLett.76.4372}{Phys.
  Rev. Lett.}} \href{http://dx.doi.org/10.1103/PhysRevLett.76.4372}{{\bf
  \bibinfo{volume}{76}}, \bibinfo{pages}{4372}}
  (\href{http://dx.doi.org/10.1103/PhysRevLett.76.4372}{\bibinfo{year}{1996}}).

\bibitem{Brandes96}
\bibinfo{author}{T.~Brandes}, \bibinfo{author}{B.~Huckestein}, and
  \bibinfo{author}{L.~Schweitzer}, \bibinfo{title}{Critical dynamics and
  multifractal exponents at the {A}nderson transition in 3d disordered
  systems},
  \bibinfo{journal}{\href{http://dx.doi.org/https://doi.org/10.1002/andp.2065080803}{Ann.
  Phys.}} \href{http://dx.doi.org/https://doi.org/10.1002/andp.2065080803}{{\bf
  \bibinfo{volume}{508}}, \bibinfo{pages}{633}}
  (\href{http://dx.doi.org/https://doi.org/10.1002/andp.2065080803}{\bibinfo{year}{1996}}).

\bibitem{Huckestein97}
\bibinfo{author}{B.~Huckestein} and \bibinfo{author}{R.~Klesse},
  \bibinfo{title}{Spatial and spectral multifractality of the local density of
  states at the mobility edge},
  \bibinfo{journal}{\href{http://dx.doi.org/10.1103/PhysRevB.55.R7303}{Phys.
  Rev. B}} \href{http://dx.doi.org/10.1103/PhysRevB.55.R7303}{{\bf
  \bibinfo{volume}{55}}, \bibinfo{pages}{R7303}}
  (\href{http://dx.doi.org/10.1103/PhysRevB.55.R7303}{\bibinfo{year}{1997}}).

\bibitem{Mantica97}
\bibinfo{author}{G.~Mantica}, \bibinfo{title}{Quantum intermittency in
  almost-periodic lattice systems derived from their spectral properties},
  \bibinfo{journal}{\href{http://dx.doi.org/https://doi.org/10.1016/S0167-2789(96)00287-4}{Phys.
  D Nonlinear Phenom.}}
  \href{http://dx.doi.org/https://doi.org/10.1016/S0167-2789(96)00287-4}{{\bf
  \bibinfo{volume}{103}}, \bibinfo{pages}{576}}
  (\href{http://dx.doi.org/https://doi.org/10.1016/S0167-2789(96)00287-4}{\bibinfo{year}{1997}}).

\bibitem{Huckestein98}
\bibinfo{author}{B.~Huckestein} and \bibinfo{author}{R.~Klesse},
  \bibinfo{title}{Diffusion and multifractality at the metal—insulator
  transition},
  \bibinfo{journal}{\href{http://dx.doi.org/10.1080/13642819808205008}{Philos.
  Mag. B}} \href{http://dx.doi.org/10.1080/13642819808205008}{{\bf
  \bibinfo{volume}{77}}, \bibinfo{pages}{1181}}
  (\href{http://dx.doi.org/10.1080/13642819808205008}{\bibinfo{year}{1998}}).

\bibitem{Huckestein99}
\bibinfo{author}{B.~Huckestein} and \bibinfo{author}{R.~Klesse},
  \bibinfo{title}{Wave-packet dynamics at the mobility edge in two- and
  three-dimensional systems},
  \bibinfo{journal}{\href{http://dx.doi.org/10.1103/PhysRevB.59.9714}{Phys.
  Rev. B}} \href{http://dx.doi.org/10.1103/PhysRevB.59.9714}{{\bf
  \bibinfo{volume}{59}}, \bibinfo{pages}{9714}}
  (\href{http://dx.doi.org/10.1103/PhysRevB.59.9714}{\bibinfo{year}{1999}}).

\bibitem{Kawarabayashi99}
\bibinfo{author}{T.~Kawarabayashi}, \bibinfo{author}{B.~Kramer}, and
  \bibinfo{author}{T.~Ohtsuki}, \bibinfo{title}{Numerical study on {A}nderson
  transitions in three-dimensional disordered systems in random magnetic
  fields},
  \bibinfo{journal}{\href{http://dx.doi.org/10.1002/andp.19995110602}{Ann.
  Phys.}} \href{http://dx.doi.org/10.1002/andp.19995110602}{{\bf
  \bibinfo{volume}{511}}, \bibinfo{pages}{487}}
  (\href{http://dx.doi.org/10.1002/andp.19995110602}{\bibinfo{year}{1999}}).

\bibitem{Guarneri99}
\bibinfo{author}{I.~Guarneri} and \bibinfo{author}{H.~Schulz-Baldes},
  \bibinfo{title}{Upper bounds for quantum dynamics governed by {J}acobi
  matrices with self-similar spectra},
  \bibinfo{journal}{\href{http://dx.doi.org/10.1142/S0129055X99000398}{Rev.
  Math. Phys.}} \href{http://dx.doi.org/10.1142/S0129055X99000398}{{\bf
  \bibinfo{volume}{11}}, \bibinfo{pages}{1249}}
  (\href{http://dx.doi.org/10.1142/S0129055X99000398}{\bibinfo{year}{1999}}).

\bibitem{Lillo00}
\bibinfo{author}{F.~Lillo} and \bibinfo{author}{R.~N. Mantegna},
  \bibinfo{title}{{Anomalous Spreading of Power-Law Quantum Wave Packets}},
  \bibinfo{journal}{\href{http://dx.doi.org/10.1103/PhysRevLett.84.1061}{Phys.
  Rev. Lett.}} \href{http://dx.doi.org/10.1103/PhysRevLett.84.1061}{{\bf
  \bibinfo{volume}{84}}, \bibinfo{pages}{1061}}
  (\href{http://dx.doi.org/10.1103/PhysRevLett.84.1061}{\bibinfo{year}{2000}}).

\bibitem{Killip01}
\bibinfo{author}{R.~Killip}, \bibinfo{author}{A.~Kiselev}, and
  \bibinfo{author}{Y.~Last}, \bibinfo{title}{Dynamical upper bounds on
  wavepacket spreading},
  \href{https://arxiv.org/abs/math/0112078}{\bibinfo{howpublished}{arXiv:math/0112078}}.

\bibitem{Yuan00}
\bibinfo{author}{H.~Q. Yuan}, \bibinfo{author}{U.~Grimm},
  \bibinfo{author}{P.~Repetowicz}, and \bibinfo{author}{M.~Schreiber},
  \bibinfo{title}{Energy spectra, wave functions, and quantum diffusion for
  quasiperiodic systems},
  \bibinfo{journal}{\href{http://dx.doi.org/10.1103/PhysRevB.62.15569}{Phys.
  Rev. B}} \href{http://dx.doi.org/10.1103/PhysRevB.62.15569}{{\bf
  \bibinfo{volume}{62}}, \bibinfo{pages}{15569}}
  (\href{http://dx.doi.org/10.1103/PhysRevB.62.15569}{\bibinfo{year}{2000}}).

\bibitem{Zhong00}
\bibinfo{author}{J.~Zhong}, \bibinfo{author}{Z.~{Z}hang},
  \bibinfo{author}{M.~Schreiber}, \bibinfo{author}{E.~W. Plummer}, and
  \bibinfo{author}{Q.~Niu}, \bibinfo{title}{Dynamical scaling properties of
  electrons in quantum systems with multifractal eigenstates},
  \href{https://arxiv.org/abs/cond-mat/0011118}{\bibinfo{howpublished}{arXiv:cond-mat/0011118}}.

\bibitem{Zhong01}
\bibinfo{author}{J.~Zhong}, \bibinfo{author}{R.~B. Diener},
  \bibinfo{author}{D.~A. Steck}, \bibinfo{author}{W.~H. Oskay},
  \bibinfo{author}{M.~G. Raizen}, \bibinfo{author}{E.~W. Plummer},
  \bibinfo{author}{Z.~Zhang}, and \bibinfo{author}{Q.~Niu},
  \bibinfo{title}{Shape of the quantum diffusion front},
  \bibinfo{journal}{\href{http://dx.doi.org/10.1103/PhysRevLett.86.2485}{Phys.
  Rev. Lett.}} \href{http://dx.doi.org/10.1103/PhysRevLett.86.2485}{{\bf
  \bibinfo{volume}{86}}, \bibinfo{pages}{2485}}
  (\href{http://dx.doi.org/10.1103/PhysRevLett.86.2485}{\bibinfo{year}{2001}}).

\bibitem{Guarneri02}
\bibinfo{author}{I.~Guarneri} and \bibinfo{author}{H.~Schulz-Baldes},
  \bibinfo{title}{Lower bounds on wave packet propagation by packing dimensions
  of spectral measures},
  \bibinfo{journal}{\href{http://dx.doi.org/10.1142/9789812777874_0001}{Math.
  Phys. Electron. J.}}
  \href{http://dx.doi.org/10.1142/9789812777874_0001}{\bibinfo{pages}{1--16}}
  (\href{http://dx.doi.org/10.1142/9789812777874_0001}{\bibinfo{year}{2002}}).

\bibitem{Cerovski05}
\bibinfo{author}{V.~Z. Cerovski}, \bibinfo{author}{M.~Schreiber}, and
  \bibinfo{author}{U.~Grimm}, \bibinfo{title}{Spectral and diffusive properties
  of silver-mean quasicrystals in one, two, and three dimensions},
  \bibinfo{journal}{\href{http://dx.doi.org/10.1103/PhysRevB.72.054203}{Phys.
  Rev. B}} \href{http://dx.doi.org/10.1103/PhysRevB.72.054203}{{\bf
  \bibinfo{volume}{72}}, \bibinfo{pages}{054203}}
  (\href{http://dx.doi.org/10.1103/PhysRevB.72.054203}{\bibinfo{year}{2005}}).

\bibitem{Damanik06}
\bibinfo{author}{D.~Damanik}, \bibinfo{title}{Quantum dynamical properties of
  quasicrystals},
  \bibinfo{journal}{\href{http://dx.doi.org/10.1080/14786430500199286}{Philos.
  Mag.}} \href{http://dx.doi.org/10.1080/14786430500199286}{{\bf
  \bibinfo{volume}{86}}, \bibinfo{pages}{883}}
  (\href{http://dx.doi.org/10.1080/14786430500199286}{\bibinfo{year}{2006}}).

\bibitem{Jitomirskaya07}
\bibinfo{author}{S.~Jitomirskaya} and \bibinfo{author}{H.~Schulz-Baldes},
  \bibinfo{title}{Upper bounds on wavepacket spreading for random {J}acobi
  matrices},
  \bibinfo{journal}{\href{http://dx.doi.org/10.1007/s00220-007-0252-0}{Commun.
  Math. Phys.}} \href{http://dx.doi.org/10.1007/s00220-007-0252-0}{{\bf
  \bibinfo{volume}{273}}, \bibinfo{pages}{601}}
  (\href{http://dx.doi.org/10.1007/s00220-007-0252-0}{\bibinfo{year}{2007}}).

\bibitem{Ng07}
\bibinfo{author}{G.~S. Ng} and \bibinfo{author}{T.~Kottos},
  \bibinfo{title}{Wavepacket dynamics of the nonlinear {H}arper model},
  \bibinfo{journal}{\href{http://dx.doi.org/10.1103/PhysRevB.75.205120}{Phys.
  Rev. B}} \href{http://dx.doi.org/10.1103/PhysRevB.75.205120}{{\bf
  \bibinfo{volume}{75}}, \bibinfo{pages}{205120}}
  (\href{http://dx.doi.org/10.1103/PhysRevB.75.205120}{\bibinfo{year}{2007}}).

\bibitem{Thiem09}
\bibinfo{author}{S.~Thiem}, \bibinfo{author}{M.~Schreiber}, and
  \bibinfo{author}{U.~Grimm}, \bibinfo{title}{Wave packet dynamics, ergodicity,
  and localization in quasiperiodic chains},
  \bibinfo{journal}{\href{http://dx.doi.org/10.1103/PhysRevB.80.214203}{Phys.
  Rev. B}} \href{http://dx.doi.org/10.1103/PhysRevB.80.214203}{{\bf
  \bibinfo{volume}{80}}, \bibinfo{pages}{214203}}
  (\href{http://dx.doi.org/10.1103/PhysRevB.80.214203}{\bibinfo{year}{2009}}).

\bibitem{Schreiber09}
\bibinfo{author}{M.~Schreiber}, \bibinfo{title}{Hierarchical diffusive
  properties of electrons in quasiperiodic chains}, {\em
  \bibinfo{booktitle}{Physics and Engineering of New Materials}\/}, edited by
  \bibinfo{editor}{D.~T. Cat}, \bibinfo{editor}{A.~Pucci}, and
  \bibinfo{editor}{K.~Wandelt}, \bibinfo{pages}{1--9}
  (\bibinfo{publisher}{Springer Berlin Heidelberg}, \bibinfo{address}{Berlin,
  Heidelberg}, \bibinfo{year}{2009}).

\bibitem{Thiem10}
\bibinfo{author}{S.~Thiem} and \bibinfo{author}{M.~Schreiber},
  \bibinfo{title}{Similarity of eigenstates in generalized labyrinth tilings},
  \bibinfo{journal}{\href{http://dx.doi.org/10.1088/1742-6596/226/1/012029}{J.
  Phys. Conf. Ser.}}
  \href{http://dx.doi.org/10.1088/1742-6596/226/1/012029}{{\bf
  \bibinfo{volume}{226}}, \bibinfo{pages}{012029}}
  (\href{http://dx.doi.org/10.1088/1742-6596/226/1/012029}{\bibinfo{year}{2010}}).

\bibitem{Thiem12}
\bibinfo{author}{S.~Thiem} and \bibinfo{author}{M.~Schreiber},
  \bibinfo{title}{Renormalization group approach for the wave packet dynamics
  in golden-mean and silver-mean labyrinth tilings},
  \bibinfo{journal}{\href{http://dx.doi.org/10.1103/PhysRevB.85.224205}{Phys.
  Rev. B}} \href{http://dx.doi.org/10.1103/PhysRevB.85.224205}{{\bf
  \bibinfo{volume}{85}}, \bibinfo{pages}{224205}}
  (\href{http://dx.doi.org/10.1103/PhysRevB.85.224205}{\bibinfo{year}{2012}}).

\bibitem{Zhang12}
\bibinfo{author}{Z.~Zhang}, \bibinfo{author}{P.~Tong},
  \bibinfo{author}{J.~Gong}, and \bibinfo{author}{B.~Li},
  \bibinfo{title}{{Quantum Hyperdiffusion in One-Dimensional Tight-Binding
  Lattices}},
  \bibinfo{journal}{\href{http://dx.doi.org/10.1103/PhysRevLett.108.070603}{Phys.
  Rev. Lett.}} \href{http://dx.doi.org/10.1103/PhysRevLett.108.070603}{{\bf
  \bibinfo{volume}{108}}, \bibinfo{pages}{070603}}
  (\href{http://dx.doi.org/10.1103/PhysRevLett.108.070603}{\bibinfo{year}{2012}}).

\bibitem{Thiem13}
\bibinfo{author}{S.~Thiem} and \bibinfo{author}{M.~Schreiber},
  \bibinfo{title}{Wavefunctions, quantum diffusion, and scaling exponents in
  golden-mean quasiperiodic tilings},
  \bibinfo{journal}{\href{http://dx.doi.org/10.1088/0953-8984/25/7/075503}{J.
  Condens. Matter Phys.}}
  \href{http://dx.doi.org/10.1088/0953-8984/25/7/075503}{{\bf
  \bibinfo{volume}{25}}, \bibinfo{pages}{075503}}
  (\href{http://dx.doi.org/10.1088/0953-8984/25/7/075503}{\bibinfo{year}{2013}}).

\bibitem{Shamis23}
\bibinfo{author}{M.~Shamis} and \bibinfo{author}{S.~Sodin},
  \bibinfo{title}{Upper bounds on quantum dynamics in arbitrary dimension},
  \bibinfo{journal}{\href{http://dx.doi.org/https://doi.org/10.1016/j.jfa.2023.110034}{J.
  Funct. Anal.}}
  \href{http://dx.doi.org/https://doi.org/10.1016/j.jfa.2023.110034}{{\bf
  \bibinfo{volume}{285}}, \bibinfo{pages}{110034}}
  (\href{http://dx.doi.org/https://doi.org/10.1016/j.jfa.2023.110034}{\bibinfo{year}{2023}}).

\bibitem{Hopjan23b}
\bibinfo{author}{M.~Hopjan} and \bibinfo{author}{L.~Vidmar},
  \bibinfo{title}{Scale-invariant critical dynamics at eigenstate transitions},
  \bibinfo{journal}{\href{http://dx.doi.org/10.1103/PhysRevResearch.5.043301}{Phys.
  Rev. Res.}} \href{http://dx.doi.org/10.1103/PhysRevResearch.5.043301}{{\bf
  \bibinfo{volume}{5}}, \bibinfo{pages}{043301}}
  (\href{http://dx.doi.org/10.1103/PhysRevResearch.5.043301}{\bibinfo{year}{2023}}).

\bibitem{Jiricek24}
\bibinfo{author}{S.~Jiricek}, \bibinfo{author}{M.~Hopjan},
  \bibinfo{author}{P.~\L{}yd\ifmmode~\dot{z}\else \.{z}\fi{}ba},
  \bibinfo{author}{F.~Heidrich-Meisner}, and \bibinfo{author}{L.~Vidmar},
  \bibinfo{title}{Critical quantum dynamics of observables at eigenstate
  transitions},
  \bibinfo{journal}{\href{http://dx.doi.org/10.1103/PhysRevB.109.205157}{Phys.
  Rev. B}} \href{http://dx.doi.org/10.1103/PhysRevB.109.205157}{{\bf
  \bibinfo{volume}{109}}, \bibinfo{pages}{205157}}
  (\href{http://dx.doi.org/10.1103/PhysRevB.109.205157}{\bibinfo{year}{2024}}).

\bibitem{Hopjan24}
\bibinfo{author}{M.~Hopjan} and \bibinfo{author}{L.~Vidmar},
  \bibinfo{title}{Survival probability, particle imbalance, and their
  relationship in quadratic models},
  \bibinfo{journal}{\href{http://dx.doi.org/10.3390/e26080656}{Entropy}}
  \href{http://dx.doi.org/10.3390/e26080656}{{\bf \bibinfo{volume}{26}}}
  (\href{http://dx.doi.org/10.3390/e26080656}{\bibinfo{year}{2024}}).

\bibitem{suntajs_prosen_21}
\bibinfo{author}{J.~\v{S}untajs}, \bibinfo{author}{T.~Prosen}, and
  \bibinfo{author}{L.~Vidmar}, \bibinfo{title}{{Spectral properties of
  three-dimensional Anderson model}},
  \bibinfo{journal}{\href{http://dx.doi.org/https://doi.org/10.1016/j.aop.2021.168469}{Ann.
  Phys. (Amsterdam)}}
  \href{http://dx.doi.org/https://doi.org/10.1016/j.aop.2021.168469}{{\bf
  \bibinfo{volume}{435}}, \bibinfo{pages}{168469}}
  (\href{http://dx.doi.org/https://doi.org/10.1016/j.aop.2021.168469}{\bibinfo{year}{2021}}).

\bibitem{Vicsek84}
\bibinfo{author}{T.~{V}icsek} and \bibinfo{author}{F.~Family},
  \bibinfo{title}{{Dynamic Scaling for Aggregation of Clusters}},
  \bibinfo{journal}{\href{http://dx.doi.org/10.1103/PhysRevLett.52.1669}{Phys.
  Rev. Lett.}} \href{http://dx.doi.org/10.1103/PhysRevLett.52.1669}{{\bf
  \bibinfo{volume}{52}}, \bibinfo{pages}{1669}}
  (\href{http://dx.doi.org/10.1103/PhysRevLett.52.1669}{\bibinfo{year}{1984}}).

\bibitem{Ohtsuki1997}
\bibinfo{author}{T.~Ohtsuki} and \bibinfo{author}{T.~Kawarabayashi},
  \bibinfo{title}{{Anomalous Diffusion at the {A}nderson Transitions}},
  \bibinfo{journal}{\href{http://dx.doi.org/10.1143/JPSJ.66.314}{J. Phys. Soc.
  Jpn.}} \href{http://dx.doi.org/10.1143/JPSJ.66.314}{{\bf
  \bibinfo{volume}{66}}, \bibinfo{pages}{314}}
  (\href{http://dx.doi.org/10.1143/JPSJ.66.314}{\bibinfo{year}{1997}}).

\bibitem{sierant_delande_20}
\bibinfo{author}{P.~Sierant}, \bibinfo{author}{D.~Delande}, and
  \bibinfo{author}{J.~Zakrzewski}, \bibinfo{title}{{Thouless Time Analysis of
  {Anderson} and Many-Body Localization Transitions}},
  \bibinfo{journal}{\href{http://dx.doi.org/10.1103/PhysRevLett.124.186601}{Phys.
  Rev. Lett.}} \href{http://dx.doi.org/10.1103/PhysRevLett.124.186601}{{\bf
  \bibinfo{volume}{124}}, \bibinfo{pages}{186601}}
  (\href{http://dx.doi.org/10.1103/PhysRevLett.124.186601}{\bibinfo{year}{2020}}).

\bibitem{Zhong98}
\bibinfo{author}{J.~X. Zhong}, \bibinfo{author}{U.~Grimm},
  \bibinfo{author}{R.~A. R\"omer}, and \bibinfo{author}{M.~Schreiber},
  \bibinfo{title}{{Level-Spacing Distributions of Planar Quasiperiodic
  Tight-Binding Models}},
  \bibinfo{journal}{\href{http://dx.doi.org/10.1103/PhysRevLett.80.3996}{Phys.
  Rev. Lett.}} \href{http://dx.doi.org/10.1103/PhysRevLett.80.3996}{{\bf
  \bibinfo{volume}{80}}, \bibinfo{pages}{3996}}
  (\href{http://dx.doi.org/10.1103/PhysRevLett.80.3996}{\bibinfo{year}{1998}}).

\bibitem{Grimm02}
\bibinfo{author}{U.~Grimm} and \bibinfo{author}{M.~Schreiber},
  \bibinfo{title}{Energy spectra and eigenstates of quasiperiodic tight-binding
  hamiltonians},
  \href{https://arxiv.org/abs/cond-mat/0212140}{\bibinfo{howpublished}{arXiv:cond-mat/0212140}}.

\bibitem{Lifshitz02}
\bibinfo{author}{R.~Lifshitz}, \bibinfo{title}{The square {F}ibonacci tiling},
  \bibinfo{journal}{\href{http://dx.doi.org/https://doi.org/10.1016/S0925-8388(02)00169-X}{J.
  Alloys Compd.}}
  \href{http://dx.doi.org/https://doi.org/10.1016/S0925-8388(02)00169-X}{{\bf
  \bibinfo{volume}{342}}, \bibinfo{pages}{186}}
  (\href{http://dx.doi.org/https://doi.org/10.1016/S0925-8388(02)00169-X}{\bibinfo{year}{2002}}).

\bibitem{Sanchez04}
\bibinfo{author}{V.~S\'anchez} and \bibinfo{author}{C.~Wang},
  \bibinfo{title}{Application of renormalization and convolution methods to the
  {K}ubo-{G}reenwood formula in multidimensional {F}ibonacci systems},
  \bibinfo{journal}{\href{http://dx.doi.org/10.1103/PhysRevB.70.144207}{Phys.
  Rev. B}} \href{http://dx.doi.org/10.1103/PhysRevB.70.144207}{{\bf
  \bibinfo{volume}{70}}, \bibinfo{pages}{144207}}
  (\href{http://dx.doi.org/10.1103/PhysRevB.70.144207}{\bibinfo{year}{2004}}).

\bibitem{Mandel08}
\bibinfo{author}{S.~Even-Dar~Mandel} and \bibinfo{author}{R.~Lifshitz},
  \bibinfo{title}{Electronic energy spectra of square and cubic {F}ibonacci
  quasicrystals},
  \bibinfo{journal}{\href{http://dx.doi.org/10.1080/14786430802070805}{Philos.
  Mag.}} \href{http://dx.doi.org/10.1080/14786430802070805}{{\bf
  \bibinfo{volume}{88}}, \bibinfo{pages}{2261}}
  (\href{http://dx.doi.org/10.1080/14786430802070805}{\bibinfo{year}{2008}}).

\bibitem{Devakul17}
\bibinfo{author}{T.~Devakul} and \bibinfo{author}{D.~A. Huse},
  \bibinfo{title}{{A}nderson localization transitions with and without random
  potentials},
  \bibinfo{journal}{\href{http://dx.doi.org/10.1103/PhysRevB.96.214201}{Phys.
  Rev. B}} \href{http://dx.doi.org/10.1103/PhysRevB.96.214201}{{\bf
  \bibinfo{volume}{96}}, \bibinfo{pages}{214201}}
  (\href{http://dx.doi.org/10.1103/PhysRevB.96.214201}{\bibinfo{year}{2017}}).

\bibitem{Jagannathan21}
\bibinfo{author}{A.~Jagannathan}, \bibinfo{title}{The {F}ibonacci quasicrystal:
  Case study of hidden dimensions and multifractality},
  \bibinfo{journal}{\href{http://dx.doi.org/10.1103/RevModPhys.93.045001}{Rev.
  Mod. Phys.}} \href{http://dx.doi.org/10.1103/RevModPhys.93.045001}{{\bf
  \bibinfo{volume}{93}}, \bibinfo{pages}{045001}}
  (\href{http://dx.doi.org/10.1103/RevModPhys.93.045001}{\bibinfo{year}{2021}}).

\bibitem{Strkalj22}
\bibinfo{author}{A.~\ifmmode~\check{S}\else \v{S}\fi{}trkalj},
  \bibinfo{author}{E.~V.~H. Doggen}, and \bibinfo{author}{C.~Castelnovo},
  \bibinfo{title}{Coexistence of localization and transport in many-body
  two-dimensional {A}ubry-{A}ndr\'e models},
  \bibinfo{journal}{\href{http://dx.doi.org/10.1103/PhysRevB.106.184209}{Phys.
  Rev. B}} \href{http://dx.doi.org/10.1103/PhysRevB.106.184209}{{\bf
  \bibinfo{volume}{106}}, \bibinfo{pages}{184209}}
  (\href{http://dx.doi.org/10.1103/PhysRevB.106.184209}{\bibinfo{year}{2022}}).

\bibitem{vega1}
\bibinfo{note}{{www.hpc-rivr.si}}.

\bibitem{vega2}
\bibinfo{note}{{eurohpc-ju.europa.eu}}.

\bibitem{vega3}
\bibinfo{note}{{ www.izum.si}}.

\end{thebibliography}

\onecolumngrid
\begin{center}
{\large \bf End matter}\\
\end{center}
\twocolumngrid

{\it Appendix A: Models with separable potentials.}
Due to separability of the potentials in the Fibonacci model, Eq.~\eqref{eq:ham}, many properties in higher dimensions contain simple relations to the 1D model.
Let us denote the eigenstates and eigenenergies of the 1D model as $|\phi_n\rangle$, and $\varepsilon_n$, respectively, with $n=1,...,L$.
Then, the eigenstates in, say, the 3D model are their products, $|\psi_{nmp}\rangle = |\phi_n\rangle |\phi_m\rangle |\phi_p\rangle$, and the eigenenergies are $E_{nmp} = \varepsilon_n + \varepsilon_m + \varepsilon_p$, with $n,m,p = 1,...,L$.

The structure of eigenstates and eigenenergies has also implications for the dynamics, since $e^{-i\hat H t}|\psi_{nmp}\rangle = e^{-i(\varepsilon_n+\varepsilon_m+\varepsilon_p)t} |\psi_{nmp}\rangle$.
For example, the transition probabilities $|\langle {\bm i}|e^{-i\hat{H}t}|{\bm i_0} \rangle|^2$, which enter the expression for the mean-squared displacement in Eq.~\eqref{def_sigma}, become products of 1D transition probabilities.
Then, one can express the mean-squared displacement in 3D as $\sigma_{\rm 3D}^2(t) = \sigma_x^2(t)+\sigma_y^2(t)+\sigma_z^2(t)$, i.e., by a sum of displacements in each direction.
Assuming an identical structure of the potentials in each direction, we replace the displacement in each direction by $\sigma_{\rm 1D}^2(t)$ and we obtain $\sigma_{\rm 3D}^2(t) = 3\sigma_{\rm 1D}^2(t)$.
More generally, in $d_l$ dimensions this relation is extended to $\sigma_{d_l {\rm D}}^2(t) = d_l\sigma_{\rm 1D}^2(t)$.

Since the mean-squared displacements in various dimensions only differ by a time-independent multiplicative prefactor $d_l$, they all give rise to the same dynamical exponent $z$, see Eq.~\eqref{def_sigma2}.
Hence, the results for $1/z$ in Fig.~\ref{fig_z} in the main text applies to all dimensions under consideration.

{\it Appendix B: Numerical tests of Eq.~\eqref{N_scaling}.}
Here, we quantitatively study the scaling $N(\epsilon)=c_N (1/\epsilon)^{d_s}$, which is given by Eq.~\eqref{N_scaling} of the main text.
For a finite system, $N(\epsilon)$ saturates to $\lim_{\epsilon\rightarrow0} N(\epsilon) \propto V$, if there are no degeneracies in the spectrum, or to lower values, if degeneracies are present. 
Examples of the function $N(\epsilon)$ are shown in Fig.~\ref{fig_ds} for the same model parameters and dimensions as in Fig.$~$\ref{fig_n}.
We note that the comparison with systems across various dimensions is more convenient if one considers $[N(\epsilon)]^{1/d_l}$, as this quantity saturates at a linear system size, $\lim_{\epsilon\rightarrow0} [N(\epsilon)]^{1/d_l} \propto L$, for all dimensions.
We observe that the scaling in Eq.~\eqref{N_scaling} describes very accurately the curves in Fig.$~$\ref{fig_ds}, and the extracted values of $d_s$ are given in each panel.

\begin{figure}[!t]
\centering
\includegraphics[width=0.98\columnwidth]{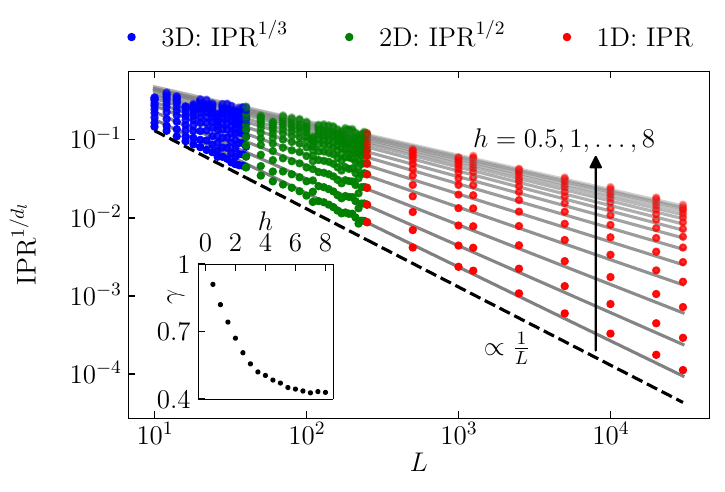}
\caption{
The $d_l$-root of the inverse participation ratio (IPR) in the corresponding $d_l$ dimensional Fibonacci models.
Results for ${\rm IPR}^{1/d_l}$ are shown on a log-log scale versus the linear system size $L$, for several values of the potential strength $h=0.5,1,\dots,8$.
The slope of the fitted curves gives the eigenstate fractal dimension $\gamma$, see Eq.~\eqref{IPR_scaling}, which is independent of $d_l$ at a fixed value of $h$.
Inset: $\gamma$ versus $h$ in the 1D model, showing that $\gamma<1$ for $h>0$.
}
\label{IPR}
\end{figure}

{\it Appendix C: Fractality of the eigenstates.}
For the 1D Fibonacci model, it is known that for any potential $h>0$ the eigenstate fractal exponent $\gamma$ satisfies $\gamma<1$~\cite{Jagannathan21}, i.e., the eigenstates are (multi)fractal. 
We calculate $\gamma$ via the averaged inverse participation ratio (IPR) that is for arbitrary dimension defined as
\begin{equation}\label{IPR_def}
    {\rm IPR} = \langle V^{-1} \sum_{\mu} \sum_{\bm i} |\langle {\bm i}|\psi_\mu \rangle|^{4}\rangle_H \;,
\end{equation}
where $|\psi_\mu\rangle$ denotes the Hamiltonian eigenstate, and $\mu=1,...,V$.
The eigenstate fractal exponent $\gamma$ is then defined via the scaling of IPR from Eq.~\eqref{IPR_def} as ${\rm IPR}\propto V^{-\gamma} = L^{-\gamma d_l}$.
In the inset of Fig.~\ref{IPR}, we plot $\gamma$ versus $h$ in the 1D Fibonacci model ($d_l=1$), and we observe that, indeed, $\gamma<1$ for all values of $h>0$.

Using separability of potentials, discussed in Appendix~A, one can show that the IPRs in higher dimensions are products of IPRs in one dimension, ${\rm IPR} = ({\rm IPR}_{\rm 1D})^{d_l}$.
From this, it follows that the eigenstate fractal dimension $\gamma$ is, at a fixed potential, identical in all dimensions, i.e., $\gamma_{\rm 1D}=\gamma_{\rm 2D}=\gamma_{\rm 3D}$.

In the main panel of Fig.~\ref{IPR}, we study the IPRs in 1D, 2D and 3D Fibonacci models for various potential values $h>0$.
For the 2D and 3D lattices it is instructive to study the corresponding $d_l$-root of the IPR, which is expected to scale as
\begin{equation}\label{IPR_scaling}
    {\rm IPR}^{1/d_l} \propto L^{\gamma} \;.
\end{equation}
This allows us to confirm in Fig.~\ref{IPR} that, indeed, $\gamma$ is identical in all dimensions.
Hence, the eigenstates of the separable Fibonacci models are fractal for all values of $h$.

\end{document}